\documentclass{aa}  

\usepackage{graphicx}
\usepackage{txfonts}
\usepackage{lipsum}
\usepackage{natbib}
\usepackage{xcolor}
\usepackage{subcaption}         
\usepackage{lscape}             
\usepackage{placeins}           
                                
\usepackage{hyperref}

\newcommand{\nooh}{\mbox{N/O$-$O/H}}

\newcommand{\tvox}{\mbox{$12 + \log(\text{O/H})$}}
\newcommand{\collisional}[3]{[\ion{#1}{#2}]\lambda#3}
\newcommand{\hbeta}{\text{H}\beta}
\newcommand{\halpha}{\text{H}\alpha}
\newcommand{\excessNO}{\mbox{$\xi_\text{N/O}$}}
\newcommand{\mstar}{\mbox{$M_\star$}}

\begin{document}

   \title{The \nooh{} relation across different large-scale environments}


   \author{Yllari K. González-Koda\inst{1}\and 
        Almudena Zurita\inst{1, 2}\and 
        Laura Sánchez-Menguiano\inst{1, 2} \and
        Ray Garner III\inst{3,4} \and 
        Sebastián F. Sánchez \inst{5,6} \and
        Bahar Bidaran \inst{1} \and
        Estrella Florido \inst{1, 2}\and
        Mónica Relaño \inst{1,2} \and
        Isabel Pérez \inst{1,2} \and 
        Simon Verley \inst{1,2} 
        }
     \institute{Universidad de Granada, Dpto. de Física Teórica y del Cosmos, Campus Fuentenueva, Edificio Mecenas, E-18071 Granada, Spain 
             \and  
             Instituto Carlos I de Física Teórica y Computacional, Facultad de Ciencias, E-18071 Granada, Spain 
             \and 
            Department of Physics and Astronomy, Texas A\&M University, 578 University Drive, College Station, TX 77843, USA     
            \and 
            George P. and Cynthia W. Mitchell Institute for Fundamental Physics \& Astronomy, Texas A\&M University, 578 University Drive, College Station, TX 77843, USA
             \and 
             Instituto de Astronom\'ia, Universidad Nacional Auton\'oma de M\'exico, A.P. 106, Ensenada 22800, BC, Mexico
             \and 
             Instituto de Astrof\'\i sica de Canarias, La Laguna, Tenerife, E-38200, Spain
             }

  \abstract
   {The large-scale structure (LSS) of the Universe shapes galaxy evolution through its intricate web-like morphology, yet its specific influence on gas-phase chemical abundances remains uncertain. Particularly, a homogeneous comparison of the \nooh{} relation, a tracer of the mass assembly and baryonic cycle, across distinct LSS environments has, until now,  been lacking. 
   In this study, we analyse a local SDSS galaxy sample curated within the framework of the CAVITY project---dedicated to studying void galaxies--- in which galaxies are classified by their large-scale structure (LSS) environment into three groups: clusters, voids, and non-cluster/non-void (NCNV) environments (historically referred to as filaments).
   We extract spectroscopic emission line fluxes from the central SDSS spectra and analyse how different galaxy properties affect the scatter in the \nooh{} relation. For that, we subtract the main \nooh{} relation from the data and study the dependence of the resulting residuals (\excessNO{}) on a set of parameters, including stellar mass (\mstar{}), $g-r$ colour, and star formation rate (SFR).  
   We find that \mstar{}$-$related variables are the best predictors of the scatter in the \nooh{} relation, with little influence of the recent star formation history of the galaxy. 
   Regarding the LSS comparison, no statistically significant differences are found for void and NCNV galaxies. Cluster galaxies, however, present a \nooh{} relation systematically lower (with $\excessNO{} \sim -0.02$).
   Consistent with this shift and with \mstar{} as a good predictor of \nooh{} deviations, we find that environmental dependences are also reflected in fundamental scaling relations: both the MZR and N/O$-$\mstar{} relations are different for cluster galaxies. These combined results point to mechanisms in the dense intracluster medium (such as strangulation or re-accretion of metal-enriched gas) as the primary drivers of these chemical offsets for galaxies in cluster environments.}

   \keywords{Galaxies: abundances --
                Galaxies: evolution --
                large-scale structure of Universe
               }

   \maketitle
   \nolinenumbers

\section{Introduction}\label{sec:introduction}

The gas-phase chemical abundances of a galaxy serve as tracers of its evolutionary stage, reflecting the complex interplay between intrinsic secular evolutionary processes and external influences~\citep[e.g.][]{Maiolino2019}. In the current paradigm, stellar and gas-phase chemical abundances are regulated by the baryonic cycle. The bare bones of the cycle are constituted by two processes: the ejection of enriched material from the stars into the interstellar medium via stellar winds and supernovae, and the subsequent formation of new stars from this enriched gas ~\citep[see e.g.][]{Matteuci2016, Weinberg2017, Curti2025}. In this context, oxygen (O) is a tracer of the chemical enrichment level of the galaxy, as it is mainly produced and expelled by massive stars on short timescales~\citep[$\lesssim~50$~Myr; see e.g][]{Rauscher2011,    Matteuchi2012, Nomoto2013, Maiolino2019}.  The production of O is considered primary, meaning it is independent of the initial metal content of the star. On the other hand, nitrogen (N) is another widely measured abundance; however, its origin is more complex, as it is produced by stars of a broad mass range (from low- to high-masses) and through both primary and secondary production. The analysis of the ratio between these two elements sheds light on the star formation history and mass assembly of galaxies \citep[e.g][]{Mallery2007, Vincenzo2016}.

The observational behaviour of the gas-phase nitrogen-to-oxygen ratio (N/O) versus oxygen-to-hydrogen (O/H), the \nooh{} relation, has been well studied and is broadly characterised by two distinct regimes in the log-log space: a plateau at low metallicity (low O/H) with high scatter, and a steep nearly linear relation between $\log\text{O/H}$ and $\log\text{N/O}$ at high O/H~\citep[see e.g.][]{Pilyugin2012, James2015}. The present-day paradigm explains this relation as the outcome of the time delay between primary and secondary nitrogen production. Whereas primary nitrogen is mainly produced by massive stars with short lifespans, secondary nitrogen is produced and released by low- and intermediate-mass (LIM) stars with slightly longer lifetimes~\citep[e.g.][]{Molla2006, Molla2010, Kobayashi2020}. In this scenario, the N/O of a galaxy is sensitive to its recent star formation history (SFH), which is typically probed by the recent star formation rate (SFR), or the specific SFR (sSFR; SFR divided by the stellar mass of the galaxy, \mstar{}). Consequently, at fixed O/H, the galaxy N/O value has been found to be driven by differences in SFH~\citep[see e.g.][]{Mallery2007, TorresPapaqui2012, Wu2013, Vincenzo2016}. An increased scatter in the \nooh{} diagram is therefore expected, induced by the diverse SFHs of individual galaxies~\citep[see also][]{Arellano-Cordova2025}.

The large-scale structure (LSS) of the Universe is a web-like distribution comprised of extremely high-density environments joined by filaments and surrounded by low-density volumes or cosmic voids~\citep[see e.g.][for reviews on LSS and galactic evolution]{Springel2006,Dayal2020}.

Many studies have assessed the correlation between galactic properties and their LSS environment~\citep[e.g.][]{Goto2003, Rojas2004, Kreckel2012, Dominguez-Gomez2023a, TorresRios2024}. Noteworthy is the effort undertaken recently by the CAVITY project~\citep{Perez2024}, which aims to study the properties of void galaxies using spatially resolved spectroscopic data~\citep[see e.g.][]{Conrado2024, Espada2026, DeDaniloff2026}. While these studies provide valuable insights into extreme environments, disentangling the complete influence of the LSS on galaxy evolution remains an active and open problem. In particular, the study of gas-phase chemical abundances has been extensive but commonly limited to O/H in the context of the $\text{O/H}-\mstar{}$ relation~\citep[MZR; e.g.][]{Cooper2008, Wu2017, Molina-Calzada2025, Bidaran2026}. 

Regarding the \nooh{} relation and the effect of the environment, previous research remains limited and has primarily focused on narrowly-defined samples, such as dwarf void galaxies~\citep{Douglass2017, Douglass2018} and high-mass cluster galaxies~\citep{Pilyugin2002}, or has lacked a clear distinction regarding the LSS~\citep[e.g.][]{Mallery2007}.
In this framework, we aim to fill this gap by characterising the \nooh{} relation variation for intermediate-high mass ($\mstar~>~10^{8.5}\;\mathrm M_\odot$) star-forming (SF) galaxies across the LSS in a self-consistent way.  To achieve this, we first conduct a general analysis of how the scatter of the N/O–O/H relationship depends on a representative set of galaxy properties. 
The aim is to identify the parameter that most effectively drives the scatter in this relation, and confirm whether the sSFR has a relevant effect on the scatter, as proposed by previous studies~\citep[e.g.][]{Mallery2007}. Later, we explore the existence of possible environmental differences in the relation based on the residuals of the \nooh{} and connect these differences to the aforementioned variable driving the scatter.

This work is structured as follows. In Sect.~\ref{sec:sample}, we describe the galaxy sample. In Sect.~\ref{sect:Methodology} we elaborate on the methodology: the full spectral fitting procedure used to extract the emission line fluxes from the observed spectra, the extinction correction, the selection criteria for SF galaxies, and the chemical abundances derivation. In Sect.~\ref{subsec:candidate_params}, we analyse the physical properties that drive the scatter in the \nooh{} relation. Finally, in Sect.~\ref{subsec:no_oh_envs} we study the effect of the LSS on \nooh{} in the context of the results, and we present the conclusions in Sect.~\ref{sec:conclusions}.

\section{Sample selection}\label{sec:sample}
The galaxy sample used here is presented in~\citet{Dominguez-Gomez2023}, and consists of 6189 cluster galaxies, 2529 void galaxies, and 15000 that are neither in clusters nor in voids (NCNV) galaxies (this definition is consistent with~\citealt{TorresRios2026,ArgudoFernandez2026} and supersedes the term `filament galaxies' in the context of the CAVITY collaboration). The LSS classification scheme is briefly summarised as follows: Void galaxies are selected to be close to the centre of their respective void by imposing a void-centric distance $r <0.8R_\text{void, eff}$, where the effective radius, $R_\text{void, eff}$, given in~\citet{Pan2012}\footnote{To maintain consistency with~\citet{Dominguez-Gomez2023} and the CAVITY parent sample definition, \citet{Pan2012} was selected rather than newer datasets such as the SDSS DR7 void catalogue by~\citet{Douglass2023}, which implements a similar methodology to~\citet{Pan2012}} defines the size (and limits) of each void. Conversely, cluster galaxies are defined as those that belong to groups with 30 or more members~\citep[a criterion based on][]{Abell1989}, $N_\text{members}\geq 30$, according to~\citet{Tempel2017}. The remaining galaxies, which comprise the intermediate density regime, $N_\text{members} < 30$ and $r \geq 0.8 \; R_\text{void, eff}$, are referred to as NCNV. This classification scheme establishes the parent sample used here and created within the CAVITY project~\citep{Perez2024}, from which a further curated subsample of void galaxies was drawn. Further details can be found in~\citet{Dominguez-Gomez2023} and~\citet{Perez2024}.

The sample, which is extracted from SDSS DR16~\citep{Ahumada2020}, comprises central spectra of galaxies obtained with a fixed 3 arcsec aperture. 
We limit the initial redshift range of the parent sample $z\in [0.01,0.05]$ to $z \in [0.025, 0.050]$ to ensure that the $\collisional{O}{II}{3727,3729}$ emission lines, employed in the estimation of the N/O and O/H abundances (see Sect.~\ref{sect:Methodology}) fall inside the spectral range covered by the SDSS spectrographs ($3800-9200$ \AA). The probed physical scales correspond to integrating up to a radius of $\sim[0.78, 1.56]$ kpc from the galaxy's centre.

Our interest focuses on SF galaxies, which are prone to inclination biases. To avoid systematic effects, we impose both an SDSS disc inclination cut of $i < 70^\circ$ and require a positive T-type from~\citet{Dominguez-Sanchez2018}; we note, however, that no noticeable changes occur when adopting lower values for the inclination limit. We further restrict the sample by imposing a signal-to-noise ratio (S/N), $\rm S/N>10$, as provided by the SDSS parameter \texttt{snMedian}, i.e. the median intensity density divided by the intensity density error in the spectra. To eliminate extreme aperture-affected cases, we discard galaxies having an SDSS Petrosian radius \texttt{petroR90\_r} > 20 arcsec, as in~\citet[][]{Dominguez-Gomez2023}. The resulting final sample comprises 1206, 890, and 4114 galaxies located in clusters, voids, and NCNV, respectively. The redshift and key physical property distributions of these sub-samples are shown in Fig.~\ref{fig:general_properties} and discussed in Sec.~\ref{sect:results}.

\section{Methodology}\label{sect:Methodology}

To ensure a consistent analysis of the \nooh{} relation, we apply full-spectral fitting to the SDSS' spectra. Through this procedure, we extract two fundamental datasets for the present analysis: (i) the emission-line fluxes, used to estimate the ionised gas chemical abundances and (ii) the SFH, from which the sSFR of each galaxy is estimated. The latter is a critical parameter for characterising the \nooh{} relation, as we reiterate, variations in the current star formation activity are known to drive significant galaxy-to-galaxy scatter (see Sect.~\ref{sec:introduction}). Analysing this and other related variables allows us to better disentangle intrinsic evolutionary effects from those imposed by the environment. We adopt a standard flat cosmology, $H_0=70$ $\rm{km}\cdot \rm{Mpc}^{-1}\rm{s}^{-1}$, $\varOmega_m = 0.3$ and $\varOmega_\varLambda=0.7$.
\subsection{Full spectral fitting}\label{subsec:fullspectral}

As a preprocessing step, we transform all SDSS spectra from vacuum to air wavelengths and correct for foreground Galactic extinction. For the latter, we use the NASA/IPAC infrared service adopting~\cite{Cardelli1989}'s extinction law and the \texttt{extinction} \texttt{python} package~\citep{Barbary2021}. The colour excess and its uncertainty are taken as the mean and standard deviation within a  5 arcmin radius circle centred on the target position, respectively, using the positions provided by ~\citet{Schlafly2011}. To convert these values to extinction, we assume the ratio $R_V= A_V/E(B-V) = 3.1$.

The full-spectral analysis is performed with a modified version of the \texttt{pyFIT3D} module~\citep{Lacerda2022}. These modifications include bug fixes in redshift estimation, improved Monte Carlo uncertainty derivations, and significant processing time optimisations.  We refer the reader to Appendix~\ref{sec:appendix_pyfit3d} and González-Koda et al. (in prep.) for further technical details.

We make use of the \texttt{maStar\_slog} SSP models, a collection of MaStar stars' empirical spectra~\citep{Yan2019} selected precisely to be used in pyPipe3D-pyFIT3D with MaNGA ~\citep{Sanchez2022}. These SSPs have been extensively benchmarked in the analysis of MaNGA, CAVITY~\citep{Sanchez2024}, and CALIFA~\citep{Sanchez2023} data. 

For the sake of completeness, we briefly discuss the pyFIT3D full-spectral fitting procedure and describe the input parameters adopted. First, the redshift ($z$), stellar velocity dispersion ($\sigma_\star$), and internal extinction ($A_V$) are determined by pseudo-randomly sampling their input search ranges with a given step. The input configurations are defined as follows: $(z, z_\text{SDSS}, z_\text{SDSS} - 0.01, z_\text{SDSS} + 0.01, 0.0005)$, $(\sigma_\star, 30, 1, 350, 10)$, and $(A_V, 0.1, 0, 2, 0.05)$, where these values correspond, in strict order, to the variable, initial guess, minimum and maximum values, and the sampling step. The value $z_\text{SDSS}$ is the spectroscopic redshift provided by SDSS DR16. These parameters are estimated by exploring the aforementioned ranges and searching for best-fit solutions, those that minimise $\chi^2$. To ensure computational efficiency, the procedure utilises a reduced set of 12 templates within a reduced spectral range of $ [3800,4700]$ \AA, where most of the spectral stellar features are found. Throughout this process, emission lines are masked in the observed spectrum.

The resulting spectrum, obtained with the best-fit parameters and the reduced number of models, works as a first-order approximation of the stellar continuum. This first fitting then allows the extraction of a gas-only spectrum by subtracting the stellar model from the observed spectrum. Next, each emission line in the gas-only spectrum is fitted with a Gaussian profile with variable central wavelength, width, and integrated flux. For the full spectral fitting, we employ the spectral range $[3800, 7100]$ \AA. In this process, we opted to fix the flux ratios of the doublet lines $F(\collisional{O}{III}{5007})/F(\collisional{O}{III}{4959})$ and $F(\collisional{N}{II}{6584})/(F\collisional{N}{II}{6548})$ to their predicted theoretical value of $\sim3$~\citep{Storey2000,Osterbrock2006}. This constraint, grounded in atomic physics, minimises the uncertainties in the line flux measurements. This assumption has been observationally validated under many different explored conditions, such as changes in redshift, reddening, or AGN emission contamination~\citep[see e.g.][]{Dimitrijevic2007,Laker2022, Natasa2025}. Finally, the gas-free spectrum (i.e. the stellar component spectrum) is fitted using a larger number of SSPs (273 models). The aforementioned procedure yields, among other properties, the measurement of all emission-line fluxes and the weighting coefficients of the stellar component, which represent the fractional contribution of each population to the total spectrum in terms of light or mass (see also Sect.~\ref{subsec:candidate_params_method}). We note that gas and stellar component fitting can be performed iteratively by adjusting the \texttt{niter} parameter. In this case, there was no significant improvement in the derived $\chi^2$ value as \texttt{niter} was increased, so it was fixed to 1.

\subsection{Emission-line selection and star-forming condition}
As mentioned in Sect.~\ref{subsec:fullspectral}, one of the primary outputs of the \texttt{pyFIT3D} pipeline is the emission-line fluxes measured from the ionised gas-only spectrum, which is obtained by subtracting the modelled stellar component (continuum and stellar absorption features) from the observed spectrum. To ensure the reliability of the measurements, we restrict the analysis to emission lines with sufficient S/N relative to the local noise in the nebular continuum (hereinafter continuum residuals).

To isolate channels containing emission lines, we use the criterion $|\lambda - \lambda_\text{cent}(l)| \leq \text{FWTM}(l) \; \forall l$, where $\lambda_\text{cent}(l)$ is the emission line central wavelength and FWTM indicates the full width at tenth maximum. To quantify the noise surrounding the line, we define the standard deviation over the continuum residuals, $\sigma_\text{res}(l)$, which is calculated over a 100 \AA{} window centred around the line wavelength, excluding channels containing the emission line ($ |\lambda- \lambda_\text{cent}(l)| > \text{FWTM}(l) \; \forall l$).

We use three different proxies for the S/N for each detected emission line: the peak intensity ($I_\text{max}$) divided by the standard deviation of the continuum ($\sigma_\text{res}$) surrounding the line, $l$,  $(\rm{S/N})_\text{peak}\equiv I_\text{max}(l)/\sigma_\text{res} (l)$, the ratio between the emission line flux ($F$) and its formal error ($\varDelta F$), $(\rm{S/N})_\text{formal}\equiv F(l)/\varDelta F(l)$, as reported by pyFIT3D (see Sect.~\ref{subsec:fullspectral}), and
\begin{equation}
(\rm{S/N})_\text{FWHM}\equiv \frac{F(l)}{\text{FWHM}(l)\cdot\sigma_\text{res}(l)},
\end{equation}
where FWHM is the full width at half maximum of the emission line. The line maximum intensity is then determined within the window of width equal to $\text{FWTM}$ centred on the line. To ensure the reliability of the measurements, we only select emission lines 
that simultaneously meet the following thresholds\footnote{
Given the fixed theoretical ratios for the $\collisional{O}{III}{4959}$, $\collisional{O}{III}{5007}$ and $\collisional{N}{II}{6548}$, $\collisional{N}{II}{6584}$ doublets, we apply the S/N threshold only to the stronger line of each pair (i.e. $\lambda5007$ and $\lambda6583$, respectively).}: $(\rm{S/N})_\text{formal}>3$, $(\rm{S/N})_\text{peak}>3$ and $(\rm{S/N})_\text{FWHM}>3$.

Next, we ensure that the emission lines originate from SF galaxies and are not affected by active galactic nuclei (AGN) emission or other ionisation mechanisms. The most commonly used diagnostic diagrams are the $\log(\collisional{O}{III}{5007}/\hbeta)- \log(\collisional{N}{II}{6583}/\halpha)$, $\log(\collisional{S}{II}{6717}/\halpha- \log(\collisional{O}{III}{5007}/\hbeta)$, and $\log(\collisional{O}{IIII}{5007}/\hbeta)- \log(\collisional{O}{I}{6300}/\halpha)$ diagrams, which classify galaxies based on a demarcation line in the flux ratios space~\citep[][BPT diagrams]{Baldwin1981, Veilleux1987}. In the analysis, we exclude the non-SF galaxies using~\citet{Kauffman2003a}'s demarcation line. We dedicate Appendix~\ref{app:composite} to assess the effect of including `composite' galaxies in the sample, those that fall between~\citet{Kewley2001} and~\citet{Kauffman2003a}'s lines.

To ensure the robustness of the ionisation source classification, we also tested alternative diagnostics such as the WHaN~\citep{CidFernandes2011} and WHaO~\citep{Sanchez2025} diagrams. Nonetheless, applying these additional criteria alongside the BPT selection reduces the sample by only $\sim 9\%$ and leaves the results in Sect.~\ref{sect:results} unchanged. This is expected, as the inclusion of $\rm{EW}(\halpha)$ primarily affects the number of retired galaxies, which are already largely excluded by the limitation to the T-Type described in Sect.~\ref{sec:sample}~\citep[see also][]{Sanchez2025}.

\subsection{Extinction correction}

To correct the emission-line fluxes for internal extinction, we derive the gas extinction $A_{V,\text{gas}}$ using the Balmer decrement, assuming case B recombination and an electronic temperature of $T=10^4$~K and density of $n_e=100 ~\rm{cm}^{-3}$. In this scenario, the ratio between the intrinsic integrated fluxes is $F_\text{intr}(\halpha)/F_\text{intr}(\hbeta$) = 2.86~\citep{Osterbrock2006}. The visual gas extinction in magnitudes can be derived as 
\begin{equation}
   A_{V,\text{gas}} = \frac{2.5}{\kappa(\lambda_{\halpha}) -\kappa(\lambda_{\hbeta})} \log_{10}\left(\frac{2.86}{F(\halpha)}{F(\hbeta)}\right),
\end{equation}
where $\kappa(\lambda)\equiv A(\lambda)/A_V = (a(\lambda) + b(\lambda)/R_V)$, is the extinction law parametrisation from \citet{Cardelli1989} \footnote{As before, $R_V=3.1$ and the \texttt{extinction} package is used~\citep{Barbary2021}.}. 
Finally, a small fraction of galaxies (less than 1\%) were excluded from the sample due to unphysical gas extinction values, $A_{V,\text{gas}}<0$.

\subsection{Chemical abundances}\label{subsec:chemical}
We employ strong-line methods to derive both N/O and O/H abundances. The list of emission-line indices and the methods (or calibrations) are summarised in Table~\ref{tab:indices} and Table~\ref{tab:calibrations}.
The N/O calibration adopted is that of N2O2 presented in~\citet{Florido2022}. Among the currently available prescriptions for local H\,{\scshape ii}  regions, to our knowledge, this calibration is based on the largest sample to date, covers the widest range of N2O2 values, and exhibits the smallest residuals. 

In the case of the O/H, it has been previously ascertained that using O calibrations that include N lines biases the results due to systematic differences in star formation efficiency across galaxies, or inside the same galaxy~\citep[see e.g.][]{Perez-Montero2009,Schaefer2020,Bresolin2025}. This is opposed to oxygen-only indices such as $R_{23}$; as a trade-off, these calibrations are bivaluated with respect to O/H. 

\begin{table}[htpb]
    \caption{Emission-line indices and the corresponding extinction-corrected line ratios used.}
    \label{tab:indices}
    \centering
    \begin{tabular}{ll}
    Index& Definition \\ \hline
    $R_{2}$ & $\collisional{O}{II}{3727,3729}/\hbeta$\\ 
    $R_3$ & $\collisional{O}{III}{4959, 5007}/\hbeta$ \\
    
    $R_{23}$ & $R_2 + R_3$ \\
    $P$ & $R_3/R_{23}$ \\
    $\rm{O}_{32}$ & $R_3/R_{2}$ \\
    O3S2 & $\left(\collisional{O}{III}{5007}/ H\beta\right)\cdot \left(\collisional{S}{II}{6717,6731}/\halpha\right)$ \\
    $\text{N2O2}$ & $\collisional{N}{II}{6583}/{\collisional{O}{II}{3727,3729}}$
    \end{tabular}
    \tablefoot{These indices are used for the derivation of O/H and N/O abundances (see also Table~\ref{tab:calibrations}).}
\end{table}

\begin{figure*}[htpb]
    \centering
    \includegraphics[width=\linewidth]{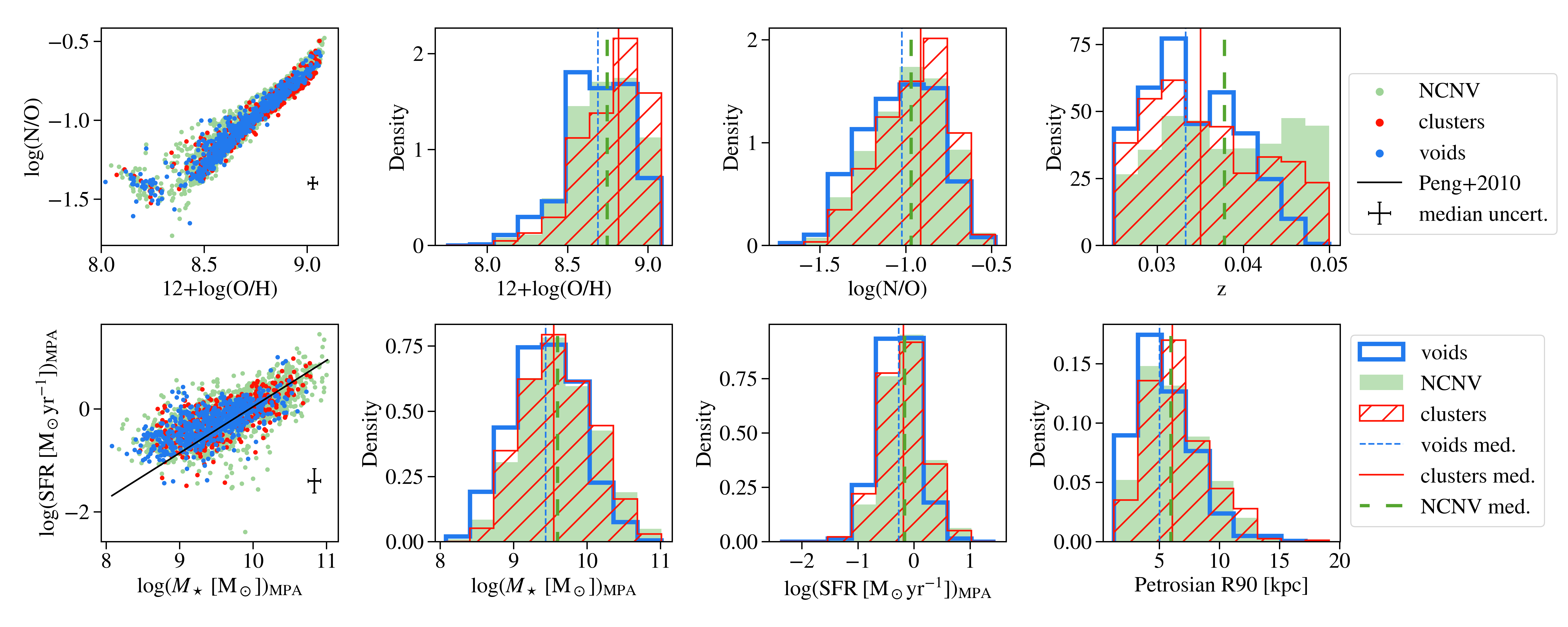}
    \caption{The top panels display the \nooh{} scatter plot, and the corresponding histograms for O/H, N/O, and redshift colour-coded depending on the environment. Similarly, the bottom panels show the SFR-\mstar{} (from MPA-JHC) scatter plot together with the main sequence from~\citep{Peng2010}, as well as the \mstar{}, SFR, and Petrosian R90 physical radius distributions. In all histograms, the vertical lines indicate the median of the respective distributions.}
    \label{fig:general_properties}
\end{figure*}

To ensure the robustness of our results, we restrict our selection of indices used to derive O/H to those based on O and sulphur emission lines.  
The emission-line index validity ranges for $R_{23}$ and $R_{23}-P$, namely C1 and C2 (Table~\ref{tab:calibrations}), are more complex than a simple interval. In the former case, C1,  the low- and high-metallicity ranges are selected following~\citet{Mallery2007}'s criterion, which uses $\log(\collisional{N}{II}{6584}/\halpha)$ and $\log(\collisional{N}{II}{6584}/\collisional{O}{II}{3727,3729})$ line ratios. In the latter, C2, we impose a simpler cut $12+\log(\text{O/H}) > 8.4$ and consistently use the high-metallicity branch formula\footnote{We consider this assumption to be reasonable, as virtually all galaxies in our sample have $\tvox{}>8.4$ (see the histogram in Fig.~\ref{fig:general_properties}, which spans the full range of metallicities derived with M91's calibration).}. For those calibrations marked with an asterisk, the relation reported in~\citet{Curti2017,Curti2020} is of the form $\log(\text{index})=f(\text{O/H})$. We have inverted them to obtain O/H and define the validity ranges (see Appendix ~\ref{app:oxygen}).  

Beyond the choice of specific indicators, the use of separate calibrations for O/H and N/O can overlook the intrinsic physical coupling between N and O enrichment. For instance, model-based O/H grids typically assume a fixed \nooh{} relationship~\citep[see e.g. Eq. 9 in][]{Mcgaugh1991}, which effectively averages out the individual impact of N abundance on the cooling efficiency of the nebula. Similarly, when the calibration is performed empirically, as when N/O is fitted as a function of N2O2, the residuals have been shown to correlate with O/H~\citep[see e.g. Fig. 2 in][]{Florido2022}.
To account for this, we incorporate into the list of adopted abundance estimators a photoionisation grid approach that obtains N/O and O/H simultaneously via Bayesian inference. This method is the one adopted in~\citet{Garner2025,Bresolin2025} and it leverages a prior set comprising  $R_{23}-\rm{O}_{32}-\rm{N2O2}$ (see Table~\ref{tab:calibrations}). Additional information regarding this methodology can be found in Appendix~\ref{app:oxygen} and in~\citet{Garner2025}.

\begin{table}[htpb]
    \caption{Calibrations of N/O and O/H and corresponding emission-line indices.}
    \label{tab:calibrations}
    \centering
    \begin{tabular}{llll}
         Ratio    & Indices                   & Range         & Refs. \\ \hline
         N/O      & N2O2                      & [-1.74, 0.62] & F22\\ 
         O/H      & $\log$(O3S2)\tablefootmark{*}             & [-0.82, 2.25] & C20 \\ 
         O/H      & $\log(R_{23})$\tablefootmark{*}           &[0.85, 0.23]   & C17 \\
         O/H      & $R_{23},\rm{O}_{32}$      &C1             & M91, K99, M07 \\
         O/H      & $R_{23},P$                &C2             & P01\\
         N/O, O/H & $R_{23},\rm{O}_{32},N2O2$ & $[-, -]$      & G25 
    \end{tabular}
    \tablefoot{List of emission-line indices, the applicability range used and the corresponding references from which the calibrations have been extracted.  Ranges C1 and C2 are more complex than a lower and an upper bound and are defined specifically for the calibration (see main text).
    \tablefoottext{*}{Indices marked with an asterisk indicate that the relation was extracted here from the formula provided in~\citet{Curti2017,Curti2020}. See the main text and Appendix~\ref{app:oxygen} for more details.}
    } \tablebib{
    (F22){~\citet{Florido2022}};
    (C20){~\citet{Curti2020}};
    (C17){~\citet{Curti2017}};
    (M91){~\citet{Mcgaugh1991}};
    (K99){~\citet{Kobulnicky1999}};
    (M07){~\citet{Mallery2007}};
    (P01){~\citet{Pilyugin2001}};
    (G25){~\citet{Garner2025}}.
    }
\end{table}

It is challenging to reliably report a `ground-truth' uncertainty for the abundance calibrations, given the many systematics and treatments among different methodologies~\citep[e.g.][]{LopezSanchez2012, Zurita2021}. For example, the uncertainties reported alongside the fitting parameters can reflect very different approximations, such as bootstrapping, Monte Carlo sampling, or the inverse of the covariance matrix. Because of this, in Sect.~\ref{sect:results} we perform the analysis with only one O/H and N/O calibration, $R_{23}-\rm{O}_{32}$ and $\rm{N2O2}$, respectively (and fixed systematic uncertainties). The main results are also corroborated in Appendix~\ref{app:oxygen} with other calibrations.

Finally, for the chemical abundances, the uncertainties used are solely the ones coming from the error reported in the emission-line flux in pyFIT3D. These were propagated using the \texttt{uncertainties} \texttt{python} package~\citep{uncertainties}.

\subsection{Candidate parameters}\label{subsec:candidate_params_method}

In this section, we present what we call `candidate parameters', a set of representative galaxy properties that trace its evolutionary stage. As stated at the beginning of this section and in Sect.~\ref{sec:introduction}, these variables will be systematically ranked and compared in Sect.~\ref{subsec:candidate_params} to find the primary drivers of the scatter in the \nooh{} relation.

Firstly, we derive the sSFR from the full spectral fitting procedure. The best-fit coefficients of the SSPs linear combination provide an approximation of the SFH in the age-metallicity space. 
Since the weights assigned to each population (characterised by age and metallicity) serve as proxies for the stellar mass formed at a specific epoch, the sSFR can be estimated by summing these weights over all metallicities within defined look-back time intervals. In particular, the recent sSFR can be inferred from the latest SFH epochs. Since the light- and mass-weighted coefficients provided for the best-fit SSPs are normalised to the total light and mass, respectively, the sSFR proxy can be calculated by summing over the weights of the youngest populations (regardless of their metallicity). We take the latest $a$ Myr and define
\begin{equation}
    \text{sSFR}_a \equiv \frac{1}{a}\sum_j^{a \; \text{Myr}}{x}_j,
\end{equation}
where we fix the summation to be over all the mass-weighted coefficients $x_j$ with SSPs younger than $a=100$ Myr. We have tested that similar results are obtained with SSPs younger than $a=20,50,75$ Myr, as the SFRs in this age range are highly correlated with each other and the SFR derived from $\halpha$~\citep[see also e.g.][]{Sanchez2022}.

Additionally, we have compiled for the galaxies in the sample other properties that may influence the \nooh{} relation, namely, SDSS $g-r$ colour, MPA-JHU's\footnote{http://www.mpa-garching.mpg.de/SDSS/DR7/} total \mstar{}, $\log M_\star$~\citep{Kauffman2003}, total SFR (derived from aperture-corrected $\text{H}\alpha$ emission line fluxes by \citealt{Brinchmann2004}), and total sSFR, $\log (\text{sSFR})_\text{MPA}$. To this set, we also include $A_{V,\text{gas}}$, as a tracer of the gas content~\citep[see e.g.][]{Gulver2009, Heiderman2010, Barrera-Ballesteros2020}, and $\halpha$ equivalent width, $\rm{EW}(\halpha)$, measured from the $\halpha$ emission-line flux and the stellar continuum from the full spectral fitting. The latter serves as a tracer of the predominance of either young or old stellar populations in the galaxy, as well as a proxy for the sSFR~\citep[see e.g.][]{Sanchez2014, Khostovan2021}. 

From the presented galaxy properties, we calculate other tracers of the galaxy evolutionary state. We define the SFR surface density within the aperture, 
\begin{equation}
\Sigma_\text{SFR, MPA} = \frac{\text{SFR}_\text{fibre, MPA}}{\pi d(z)^2},
\end{equation}
where $d(z)$ corresponds to the physical radius covered by SDSS's aperture\footnote{Total and fibre SFR correspond to the aperture-corrected and non-aperture-corrected measurements.}. Lastly, to probe the gravitational potential, we include the ratio $\varPhi \equiv \log (\mstar / R_{\rm{eff}})$, where $R_\text{eff}$ corresponds to the effective radius calculated using the Petrosian R50 and R90 in the $r$-band as provided by SDSS and Eq. 6 of~\citet{Graham2005}. 

\section{Results and discussion}\label{sect:results}
To disentangle the effect of the environment on the \nooh{} relation, it is essential to identify the physical variables that drive the scatter. Systematic differences in this relation between galaxies inhabiting different environments would most probably be traced back to such variables. Following this reasoning, the section is structured as follows: First, we characterise the overall properties of the sample. Then, in Sect.~\ref{subsec:candidate_params}, we investigate which variables drive the deviations in the \nooh{} plane with respect to the principal relation (secondary dependence with the candidate parameters introduced in Sect.~\ref{subsec:candidate_params_method}). Finally, in Sect.~\ref{subsec:no_oh_envs} we investigate the deviations in the \nooh{} relationship for galaxies residing in distinct environments (voids, NCNV, and clusters) and discuss them in the context of our previous findings.

The applied cuts in the S/N of the emission lines involved in the employed calibrator to estimate gas-phase chemical abundances determine the number of galaxies in the sample. In this case, they amount to 414, 2284, and 610 for the subsamples of voids, NCNV, and clusters, respectively.

In Fig.~\ref{fig:general_properties} we characterise the sample by showing the \nooh{} and $\text{SFR}-M_\star$ relations, together with the histograms of all involved variables for galaxies in different LSS environments. Additionally, we present redshift and physical Petrosian radius histograms\footnote{Here 'physical' refers to the Petrosian R90 radius transformed to linear distance via the luminosity distance derived from the spectroscopic redshift. For cluster galaxies, spectroscopic redshift is not a good estimator of luminosity distance. Nonetheless, the symmetry of these peculiar velocities along the line of sight should increase the dispersion without systematically biasing the properties averaged over the entire cluster population. }. At first sight, all subsamples exhibit similar distributions in both two-dimensional planes: the $\text{SFR}-\text{\mstar{}}$ (star-forming main sequence, SFMS) and the \nooh{} plane.  This is because all of the galaxies are, by definition, in the narrow range of values imposed by the SF condition; therefore, the resulting sample is comprised of galaxies in (or close to) the blue cloud~\citep[see e.g.][]{Strateva2001, Baldry2004}. Any notable differences among the sub-samples, such as the distinct $z$, O/H, and \mstar{} distributions, come directly from the parent sample, reflecting a combination of intrinsic environmental differences and survey selection biases. Additionally, we note that the \nooh{} relation shows a knee at $\tvox~\sim~8.4$ as one would expect from the low-O/H regime of the \nooh{} relation. Nonetheless, we advise some caution about these (sparse) datapoints, as this O/H regime corresponds as well to the degeneracy region between the metal-rich and metal-poor branches characteristic of the used O/H calibration, where metallicity estimates may be more uncertain.  Lastly, the deviations from~\citet{Peng2010}'s SFMS at the low-\mstar{} end are induced by completeness effects, where bluer galaxies are preferentially detected over redder (and fainter at fixed \mstar{}) ones~\citep[see e.g.][]{Kaviraj2025,TorresRios2026}. This effect is more pronounced below $\log \mstar [M_\odot] \lesssim 9.5$.

\subsection{Explaining the scatter in the N/O--O/H relation}\label{subsec:candidate_params}

In order to explore the possible existence of a secondary dependence of the \nooh{} relation on any of the candidate parameters, we implement a two-step analytical framework. First, we extract the primary \nooh{} relation, and then we examine the residuals of N/O as a function of the candidate parameters. This framework, referred to two arbitrary variables, $u$ and $x$, reads as follows:
\begin{itemize}
    \item[i)] Define $N$ equally spaced bins in the range of the minimum and maximum values of $u$.
    
    \item[ii)] For each bin, $i$, calculate both the median in the observed $u$ and $x$ variables, $(u,x)_\text{median, $i$}$.
    
    \item [iii)] Perform a cubic spline interpolation over all medians $(u, x)_\text{median, $i$}$, defined as $x_\text{interp}(u)$.
    
    \item [iv)] Calculate the residual of the variable, $\delta x(u)$, as the $x$-distance to the interpolation $\delta x(u)= x - x_\text{interp}(u)$.
\end{itemize}

We set the reference variable, in which the binning is performed, to $u= \,$\tvox{} with $N=12$ bins\footnote{This value is arbitrary; we have checked that when considering reasonably large values of bins ($N >8$), the results do not significantly vary (more so in the linear regime, where most of the galaxies in the sample reside). 
Additionally, to apply this methodology, one should ensure that the relation between the variables is monotonic and invertible (e.g. with no plateaux).}. We then apply this procedure to all candidate parameters and to $\log(\text{N/O})$, yielding a set of new variables decoupled with respect to O/H.  
We note that this procedure primarily serves to better isolate the underlying physical correlations. Indeed, although we find consistent trends when investigating the relation of the candidate parameters themselves with the excess in N/O, $\delta \log(\text{N/O})$ (hereafter \excessNO{}, similarly to ~\citealt{Schaefer2020}), the use of residuals effectively removes the primary dependence on metallicity, thereby enhancing the strength and significance of the secondary correlations.
All residuals are calculated relative to the trend of the entire sample (comprising voids, NCNV, and clusters).

\begin{figure}
    \centering
    \includegraphics[width=\linewidth]{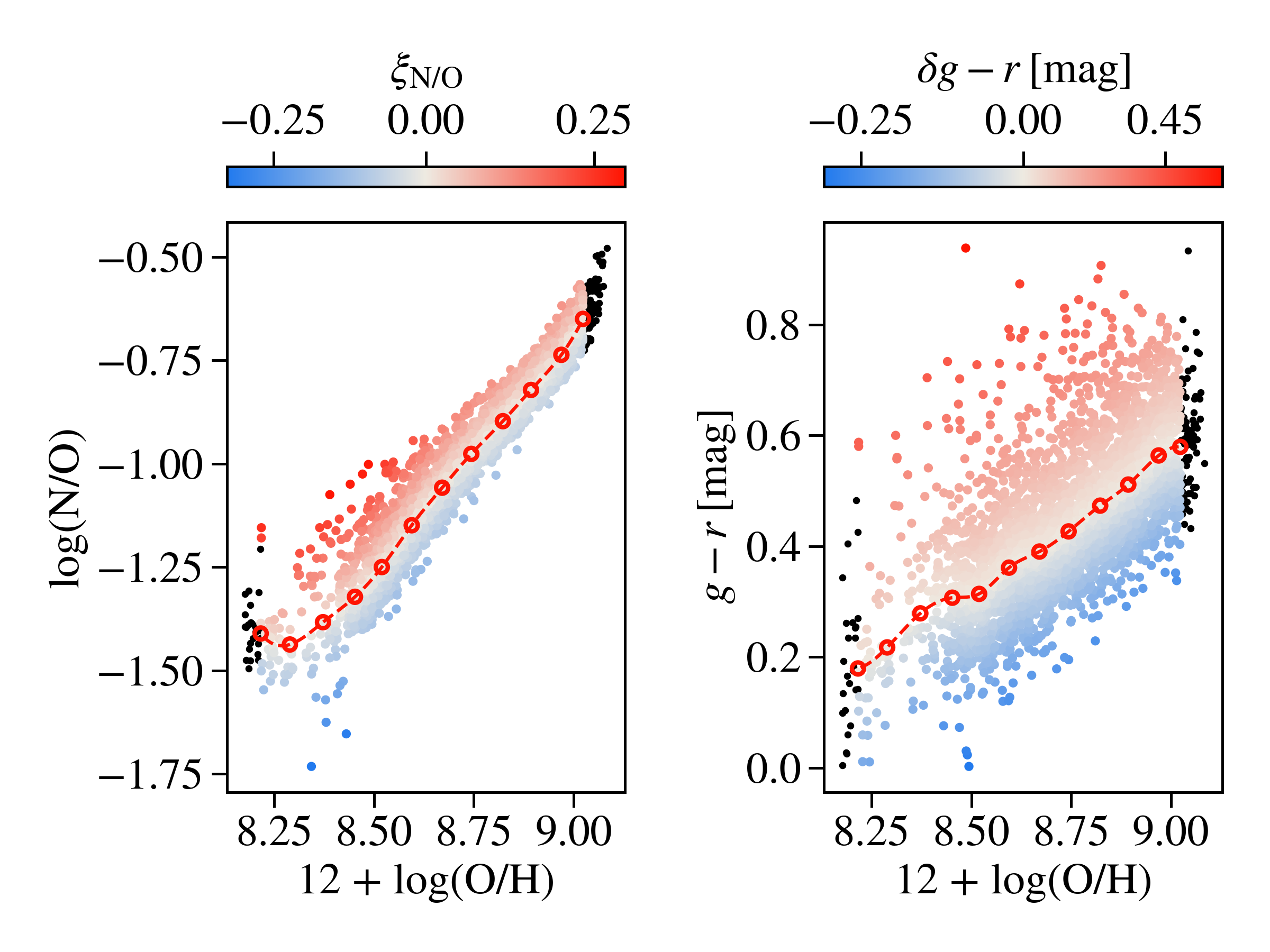}
    \caption{
    Results of the procedure described in Sect.~\ref{subsec:candidate_params}. The red dashed line represents the cubic spline interpolation applied to the binned medians (empty red markers). The data points are colour-coded following their corresponding residual, $\delta x = x - x_\text{interp}$.
    Data points with O/H lower (greater) than the lowest (highest) O/H binned median are shown in black. In the left panel, N/O versus O/H along with the residual N/O, \excessNO{}. In the right panel, an example of the same procedure applied to a candidate parameter, the $g-r$ colour.
    }
    \label{fig:excess_definitions}
\end{figure}

In Fig.~\ref{fig:excess_definitions}, we illustrate the procedure applied to \excessNO{}\, and $\delta g - r$, where we show the median values calculated in each bin along with the resulting cubic spline fit. 
To avoid extrapolation, we restrict the analysis of the residuals to the interpolated range given by the minimum and maximum \tvox{}-medians (black markers in Fig.~\ref{fig:excess_definitions}). Moreover, we do not include in this and future analyses the sparse low-oxygen-abundance galaxies by imposing a lower limit of $8.17$, corresponding to the 1st percentile of the \tvox{} distribution. This criterion restricts the analysis to the O/H range where we have enough galaxies to ascertain statistically significant results. Finally, we draw attention to the fact that, for \excessNO{}, since the interpolation is almost linear in the given O/H range, taking either the orthogonal distance, $d_\perp$, to the curve or the vertical distance would yield similar results, as they are related for a straight line via $d_\perp= \frac{1}{2}\xi_\text{N/O} \sqrt{1 + m^2}$, where $m$ is the slope of the linear relation.

Now, we explore which of the candidate parameters introduced in Sect.~\ref{subsec:candidate_params_method} (hereafter denoted as $x$, and their respective residuals, $\delta x$) is best to describe the deviations from the principal \nooh{} relation, namely \excessNO{}. These parameters are expected to correlate with all the underlying processes responsible for the scatter in the \nooh{} plane, therefore providing the context to interpret the results. 

To quantify the correlation between residual variables $\delta x$ and $\xi_\text{N/O}$, we rank the candidate parameters based on their median Spearman correlation coefficient ($\rho$). The coefficients are derived from taking $n=100$ Monte Carlo realisations of the dataset\footnote{This approach is adopted for easy reproducibility, although we have checked that similar results are obtained with more sophisticated algorithms, such as random forests.} $\{(\text{\tvox{}}_i, x_i, \log\text{(N/O)}_i)\}_{j=1}^{n}$, where the same $i$ indicates that the data is paired as a tuple. For each realisation $j=1,\dots, n$, every data point is modelled as a Gaussian distribution centred on its nominal value with a standard deviation equal to its uncertainty. Notwithstanding, this approach does not account for the intrinsic cross-correlation between variables, such as O/H and N/O. However, this approximation is valid for our results as it will only reduce the correlation strength without introducing spurious results. Thus, one must consider these values as conservative lower estimates\footnote{There is another caveat about this approach, as it relies on uncertainties coming from different works, which are prone to systematic differences. Because of this, we have also used bootstrapping instead for value and uncertainty estimation, leading to higher Spearman coefficients but preserving the ranking of the candidate parameters. Additionally, to take into account the cross-correlation in the \nooh{} relation, we have tested applying the Monte Carlo approach directly to the emission line fluxes and their errors and found no significant change in the results.}.
\begin{figure*}
    \centering
    \includegraphics[width=1\linewidth]{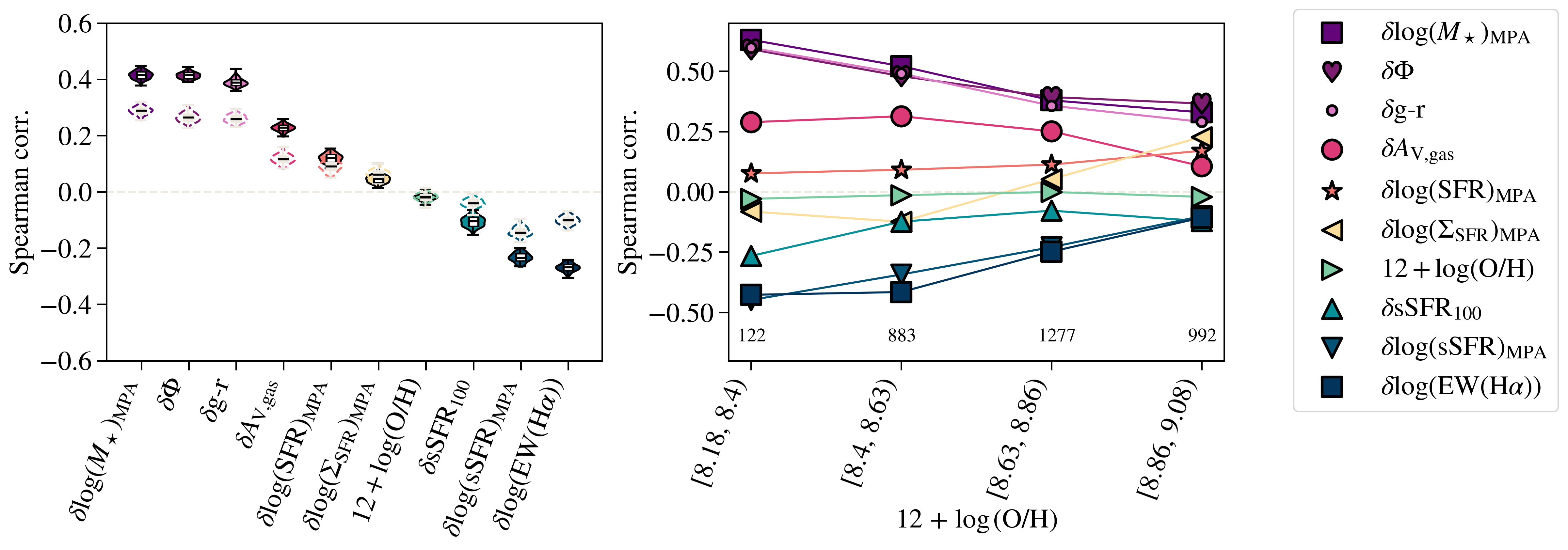}
    \caption{The left panel shows the distribution of the Spearman correlation coefficients between the residuals of the candidate variables on the x-axis and the residual N/O for the Monte Carlo sampling. The list comprises, from left to right: total \mstar{}, baryonic gravitational potential traced by $\varPhi=\log(\mstar/R_\text{eff})$, colour $g-r$, gas extinction derived from the Balmer decrement, total SFR, SFR surface density inside the fibre, O/H, sSFR derived from pyPipe3D's SFH, the ratio between the total SFR and the total \mstar{} from the MPA-JHU catalogue, and $\halpha$ equivalent width. The boxes in the plots show the 25th, 50th, and 75th percentiles, whereas the violins encapsulate the whole distribution. While the filled violin plots and white boxes correspond to the sample used throughout the main text, the dashed violin plots and grey boxes correspond to the mock sample defined in Appendix~\ref{sec:app_pearson}. Note also that \tvox{} is not defined as a distance, $\delta$ (see Appendix~\ref{app:biases}). The right panel shows the medians of the Monte Carlo Spearman coefficient distributions per O/H bins for the same variables. Bottom numbers indicate the galaxy count per bin.}
    \label{fig:ssfr_indic}
\end{figure*}

In Fig.~\ref{fig:ssfr_indic}, we show the ranking of the residual candidate parameters, $\delta x$, based on their Spearman correlation coefficients with respect to \excessNO{}. The most significant variables, i.e. those that better explain the scatter in the \nooh{} relation, are the total \mstar{} and $\varPhi$ ($\rho\sim 0.42$), together with the colour, $g-r$ ($\rho\sim 0.38$), and followed by, EW($\halpha$) $(\rho\sim-0.27)$, the sSFR from MPA-JHU catalogue ($\rho\sim-0.24)$, and $A_{V,\text{gas}}$, ($\rho \sim 0.23$). In contrast, both the aperture-corrected SFR and the fibre-based SFR density $(\Sigma_\text{SFR})$ show almost no correlation with \excessNO{}\;($\rho\sim 0.1$), indicating that they are not primary drivers of the observed scatter. 

One caveat about this approach is that the O/H dependence is lost in the analysis. Galaxies with different O/H (or \mstar{}, given the MZR) may be affected differently by secular or external processes, meaning that different explanatory variables would be relevant in different O/H (or \mstar{}) ranges. To account for this, we divide the O/H range into four equal bins with boundaries defined at \tvox{}~=~$[8.18, 8.40, 8.63, 8.86, 9.08]$ and repeat the same analysis within each bin. We show the result in the right panel of Fig.~\ref{fig:ssfr_indic}. We find that the Spearman coefficients for the most highly correlated variables decrease with increasing O/H, possibly indicating an increasing dominance of decorrelated noise for more metal-rich galaxies. Specifically, if the total scatter stems from a combination of both decorrelated noise and physical-observational scatter, a decrease in correlation implies a change in the relative contribution of the two sources. In this case, it would be plausible that galaxies at the high-metallicity end (and high-mass end, where the MZR plateaus) were less affected by external processes --such as gas inflows or galaxy-galaxy gravitational interactions-- thus exhibiting lower `physical' scatter in the \nooh{} plane. Conversely, variables with low global correlations, such as SFR, which remain statistically insignificant across most bins, are likely dominated by random noise. 

From these results, it follows that the best tracers of $\xi_\text{N/O}$ are those variables associated with the galaxy's colour, \mstar{}, and $\varPhi$. We have tested that this remains valid even when considering galaxies in voids, NCNV, or clusters independently, with only minor variations in the coefficients. It is also worth noting that, since these galaxies are SF, colour and \mstar{} are expected to be correlated (they populate the blue cloud, ~\citealt{Strateva2001,Baldry2004}). The positive Spearman coefficients for \mstar{} and $g-r$ colour point to the fact that redder, more massive galaxies tend to have larger N/O at fixed O/H. More importantly, these variables show a stronger correlation than those related to the SFR, including also the sSFR, both from the full spectral fitting procedure and the MPA-JHU's estimation (see Sect.~\ref{sect:Methodology}). In the intermediate regime, the correlation with $A_{V,\text{gas}}$ can be naturally explained by the $A_{V,\text{gas}}-\mstar{}$ relation~\citep[see e.g.][]{Garn2010}. Consistently, we obtain similar results when analysing the residuals of the candidate parameters relative to \mstar{} ($x-\mstar{}$ relation) instead of \tvox{} (see Appendix ~\ref{sec:app_pearson}).

\begin{figure*}
    \centering
    \includegraphics[width=0.8\linewidth]{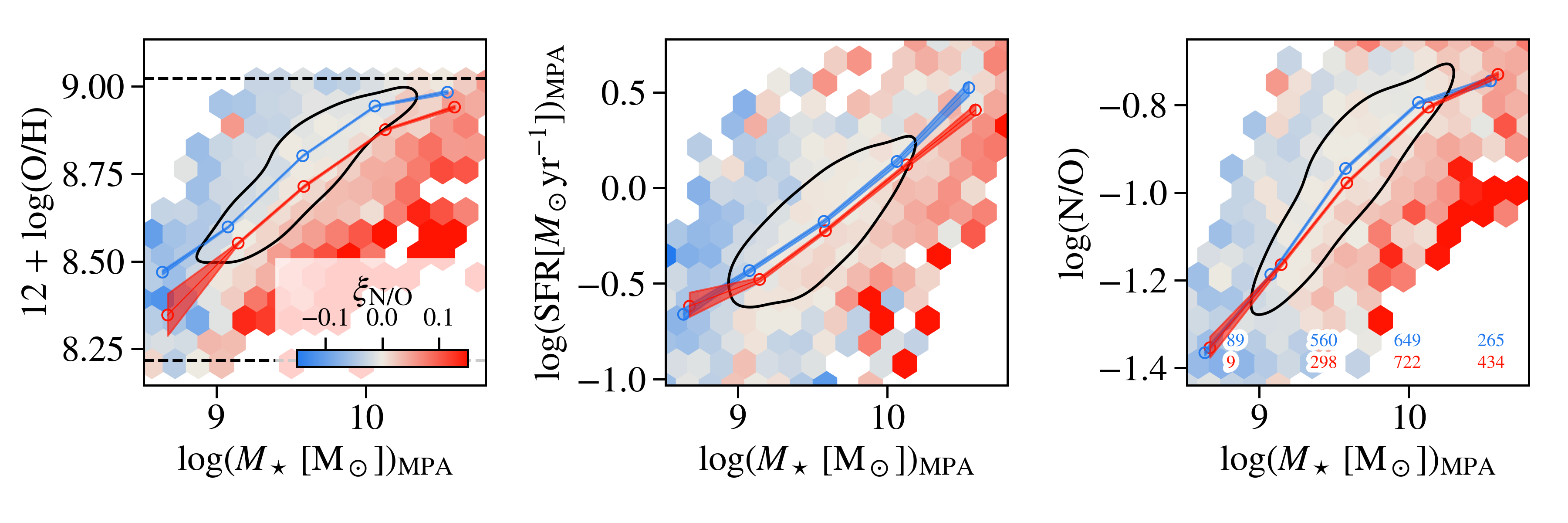}
    \caption{
    From left to right: the mass-metallicity relation (MZR), star-forming main sequence (SFMS), and N/O-\mstar{} relation. The hexagonal bins are colour-coded according to the median \excessNO{} value calculated for the galaxies inside each bin, with the colour bar saturated to the range \excessNO{} $\in[-0.15, 0.15]$ for visual clarity. Contours indicate the 50th percentile of the galaxy density distribution. 
    Red (blue) markers show the median values per mass bin for the \excessNO{}$\geq0$ (\excessNO{}$<0$) subpopulations, while the corresponding shaded lines indicate their 25th--75th percentile ranges for the Monte Carlo sampling. In the rightmost panel, the numbers indicate the galaxy counts within each \mstar{} bin for the positive and negative $\xi_\text{N/O}$. The black dashed lines indicate the boundaries of the interpolation performed to calculate \excessNO{}.}
    \label{fig:main_relations}
\end{figure*}

Motivated by the previous analysis, in Fig.~\ref{fig:main_relations} we show \mstar{} as the tracer of the galaxy evolutionary stage on the x-axis versus other relevant properties, namely, O/H, SFR, and N/O (\mstar{} scaling relations).
Given that $\delta x$ is defined as the vertical distance between the observed value and the interpolated relation with O/H, the strong correlation found between $\delta\text{\mstar{}}$ and $\xi_\text{N/O}$ directly links the variations in the \nooh{} relation with those of the (inverse) MZR.  In the left panel of Fig.~\ref{fig:main_relations}, we show the MZR with the sample divided into galaxies presenting excess (\excessNO{}$>0$, red) and deficiency (\excessNO{}$<0$, blue) of N/O. When a galaxy is situated below the median MZR, it is most probably found above the median baseline of the \nooh{} relation; that is, galaxies presenting an excess (deficiency) of N/O tend to be more metal-poor (metal-rich) than the median population at fixed \mstar{}. 

Something similar occurs for the SFR, but to a lesser extent due to the weaker correlation found between $\delta\text{SFR}$ and \excessNO{}. As shown in the middle panel of Fig.~\ref{fig:main_relations}, there seems to be a connection between variations in the \nooh{} relation and in the SFMS. Galaxies with positive (negative) \excessNO{} are slightly below (above) the median SFMS. Additionally, the same trend found in the \nooh{} plane also appears in the N/O-\mstar{} relation but is significantly less pronounced, as shown in the right panel of Fig.~\ref{fig:main_relations} by the small separation between the positive and negative \excessNO{} sequences.

Although we have adopted a \mstar-fixed approach (binning performed in this axis) in the figure, we can extract important conclusions from the colour-coded bins: the maximum separation between \excessNO{} sequences occurs approximately at a fixed O/H, SFR, or N/O, extending primarily along the \mstar{} axis. Specifically, the direction of the gradient indicates that higher (lower) \mstar{} at fixed SFR, O/H, or N/O correlates with higher (lower) values of \excessNO{}. This behaviour reinforces the role of \mstar{} as the primary mediator of the physical processes that underlie the variations of \excessNO{}.

These findings align with the existing literature. \citet{Belfiore2017}, ~\citet{Schaefer2020}, and~\citet{Boardman2024a} reported a significant correlation between \mstar{} and the scatter in the \nooh{} relation using MaNGA data. More recently, using machine-learning techniques including random forests and neural networks,~\citet{Curti2022} disentangled the drivers behind a galaxy's position in the $[\ion{N}{II}]$ BPT diagram. They found that the best explanatory variables for the offset from the main locus in the diagram (after other line ratios involving $[\ion{N}{II}]$ and used effectively as proxies of the N/O abundance) were mass, colour, and $\rm{EW}(\halpha)$  (see Fig. 4 therein). Due to the correlation between the position in the \nooh{} plane and the BPT diagram, these results are also in agreement with our findings (as shown in their Fig. 8).

Regarding the studies that have systematically addressed the secondary dependences of the \nooh{} relation with galactic properties, most have focused on the sSFR and the SFH as the main drivers. In~\citet{Mallery2007}, they analysed a sample of UV-selected starburst galaxies to assert the dependence of the offsets in the \nooh{} relation on the sSFR. They find a strong anticorrelation consistent with our results and briefly discuss the dependence of N/O residual on \mstar{} and colour~\footnote{We briefly note that in~\citet{Mallery2007}, the analysis is performed following a division into bins without eliminating the primary relation. When this approach is adopted, we ask for caution when assessing secondary dependencies, as it may fail to fully decouple the explanatory variable (in this case, sSFR) from the O/H. As a consequence, binning may introduce spurious correlations stemming from the dependence between O/H and the secondary variable, as we have confirmed through our own testing.}.

~\citet{TorresPapaqui2012} studied the differences between nitrogen-enriched selected SF galaxies and `regular' SF ones using STARLIGHT~\citep{CidFernandes2005} to derive SFHs. Their discussion focuses on variations in SFH, attributing the highest N/O excesses to changes in the recent SFH (analogous to sSFR) and extreme SFR surface densities. Although we do not find this connection with SFR or $\Sigma_\text{SFR}$, they also report a strong correlation with bulge stellar mass and the $B$-band magnitude, $M_B$, consistent with our findings (given that both parameters scale with  \mstar). 
These authors attributed the trends with mass-related quantities to the high efficiency of differential winds during the early-time starburst episodes, occurring when the galaxies were less evolved and possessed shallower gravitational potential wells, consistent with the correlation found with $\varPhi$.

As reviewed here, the focus when analysing the scatter in the \nooh{} relation has been largely positioned in the sSFR, attributing the observed variations in the SFR relative to a given \mstar{}. 
This analysis does not disprove nor contradict these ideas; however, it shows that applying a broader and systematic framework to explore all possible correlations reveals that the \mstar{} dependence is the predominant one (even more so than that hinted by~\citealt{TorresPapaqui2012}). Contrary to expectation, SFR does not seem to carry relevant information about the offsets from the primary \nooh{} relation, and the found correlation with sSFR by previous works may be the subproduct of the cross-correlation between sSFR and \mstar{} (by definition $\log\text{sSFR}\equiv \log\text{SFR} - \log\text{\mstar{}}$).
This phenomenon can be explained by the combination of both physical and non-physical effects: either \mstar{} is physically a better tracer of the dominant factors at play in the scatter of the \nooh{} relation (e.g. galactic winds), or the \mstar{} determination is, in general, more robustly determined than the SFR derived either from $\halpha$ or full spectral fitting. These conclusions are, of course, subject to the specific timescales used for the SFR derivation (around 50 Myr)\footnote{Moreover, our attempts to identify an alternative SFH tracer across different timescales from the SSP coefficients, which could correlate better with $\xi_\text{N/O}$ than \mstar{}, were unsuccessful.}.

In the current paradigm of galactic evolution, gas-phase chemical abundances are dominated by secular evolution and modulated by the interplay between gas inflows and outflows~\citep[see e.g.][]{Finlator2008, Vincenzo2016, Curti2025}. Regarding the scatter in the \nooh{}, as well as the MZR, the most relevant factors at play are considered to be differential galactic winds~\citep[see e.g.][]{TorresPapaqui2012,Vincenzo2016}, in which core-collapse products such as O are preferentially expelled to the intergalactic medium by kinetic energy; and metal-poor gas accretion~\citep[e.g][]{SanchezAlmeida2014}. 

Interpreting our findings within this galactic evolution framework, these results would suggest that \mstar{} is a good predictor of the net influence of these phenomena on the observed chemical content. Moreover, the found connection between the deviations in the \nooh{} relation and the MZR implies that the dominant (or net) effect is such that, at a given \mstar{}, galaxies lying above the median MZR fall below the median \nooh{}. This correlation can only be explained if, at fixed \mstar{}, the increase (decrease) in O/H was dominant over any possible increase (decrease) in N/O. In fact, what we observe is that at fixed \mstar{}, the \excessNO{} division is clear in the MZR but less pronounced in the N/O-\mstar{} relation. This anticorrelation is compatible with the aforementioned mechanisms for the scatter in the \nooh{} relation. For example, if differential galactic winds driven by core-collapse supernovae were at play, oxygen would be preferentially subtracted via the high-energy winds induced by the death of massive stars; consequently, a decrease in O/H and an increase in N/O take place. Similarly, if pristine gas were to be accreted, we would see a decrease in O/H at fixed N/O; the net effect is that the galaxy is shifted above the primary \nooh{} relation, thus showing a positive \excessNO{}.

From the previous analysis, it is clear that at fixed \mstar{}, the differences in \excessNO{} are mostly correlated with variations in the MZR. While the N/O$-\mstar{}$ relation remains largely invariant under the \excessNO{}\, classification (Fig.~\ref{fig:main_relations}, right panel), the MZR exhibits a clear connection with \excessNO{}, across all mass ranges. This means that the variance in \excessNO{} is dominated by fluctuations in the MZR at a fixed \mstar{}, with little to no contribution from variations in the N/O$-\mstar{}$ relation. Accordingly, we suggest that a possible explanation for why mass-related quantities act as primary predictors of \excessNO{} is that the N/O$-\mstar{}$ relation is less sensitive to gas inflows and outflows that affect both the MZR and the \nooh{} relation, as discussed, e.g., in~\citet {Masters2016} and~\citet{Boardman2024}. Under this hypothesis, variations in the MZR would be the main drivers of the deviations in the \nooh{} relation, consistent with our findings.

To validate that the reported trends are driven by underlying physics rather than systematic biases, in Appendices~\ref{app:biases} and~\ref{sec:app_pearson} we identify and quantify separate sources of non-physical effects to ensure the reliability of the conclusions. Among the examined factors, we include: potential biases in the \excessNO{} definition, cross-correlation among variables, effects arising from the S/N selection criteria, aperture coverage effects, and the choice of oxygen abundance index and calibration. However, we acknowledge that the use of integrated spectra naturally averages over complexes of unresolved $\ion{H}{II}$ regions and secondary ionisation sources within the aperture, an effect that cannot be fully disentangled with single-fibre data alone, despite our multi-diagram diagnostic testing (Sect.~\ref{sect:Methodology}).

\subsection{\nooh{} relation and its connection to \mstar{} scaling relations across LSS environments}\label{subsec:no_oh_envs}

\begin{figure}
    \centering
    \includegraphics[width=\linewidth]{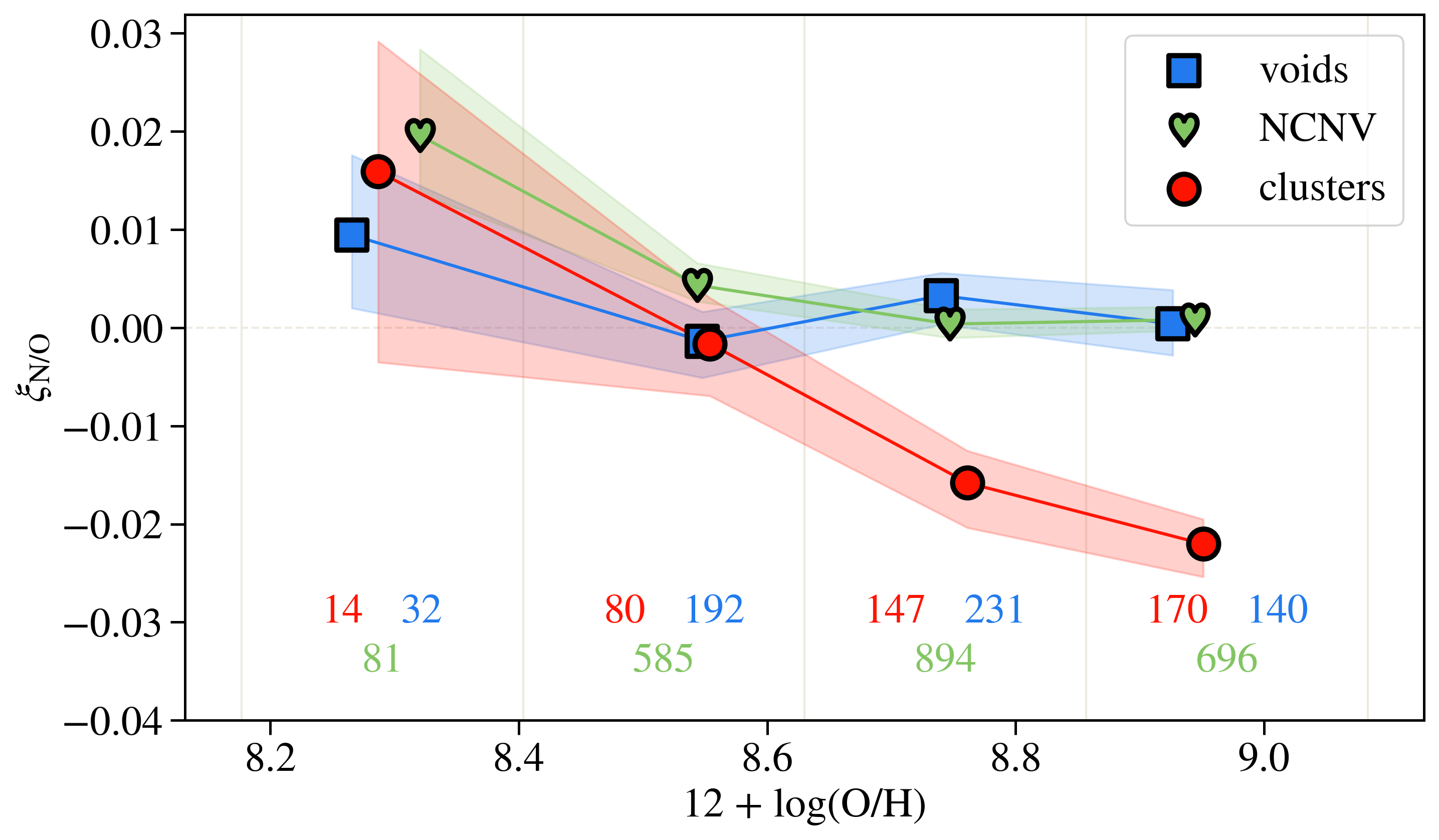}
    \caption{Median residual N/O, $\xi_\text{N/O}$, versus $12+\log(\text{O/H})$ in bins of 0.22 dex for different environments. The markers and shaded regions represent the median and the 25th-75th percentiles of the Monte Carlo distributions, respectively. Markers are centered at the median $12+\log(\text{O/H})$ and $\xi_\text{N/O}$ values within each bin. Vertical dashed lines show the bin edges, and the integers at the bottom of the panel indicate the number of galaxies per bin for cluster, NCNV, and void environments.}
    \label{fig:no_excess_envs}
\end{figure}

Having identified the primary parameters driving $\xi_\text{N/O}$, we now examine the \nooh{} relation across different environments. We present in Fig.~\ref{fig:no_excess_envs} $\xi_\text{N/O}$ as a function of $12+\log(\text{O/H})$ for galaxies residing in clusters, NCNV, and voids. The residuals are calculated relative to the global spline fit derived for the whole sample (i.e. the one calculated in Sect.~\ref{subsec:candidate_params}). We employed the Monte Carlo procedure introduced in Sect.~\ref{subsec:candidate_params} to estimate the uncertainties in the median  $\xi_\text{N/O}$ and \tvox{} values. 

First, we note that galaxies in NCNV constitute the great majority of the sample and, as such, serve as the reference baseline against which other environments are compared. This is evident in the figure, where NCNV galaxies exhibit \excessNO{}$\sim 0$ across all bins except in the first one, in which the number of galaxies is comparable to both voids and clusters\footnote{We remark that the low-limit, \tvox$~<~8.5$, is to be treated with caution, as the $R_{23}-[\tvox{}]$ degeneracy is more pronounced here.}.
In the case of voids, no significant differences with respect to the NCNV sample are observed. The largest discrepancies are found in the lowest O/H bins, but with a substantial overlap of the error bars. In contrast, cluster galaxies exhibit systematically N/O deficiency (up to \excessNO{}$=-0.025$) in the high-metallicity end (\tvox{}$>8.63$). 

\begin{figure*}
    \sidecaption
    \centering
    \includegraphics[width=0.7\linewidth]{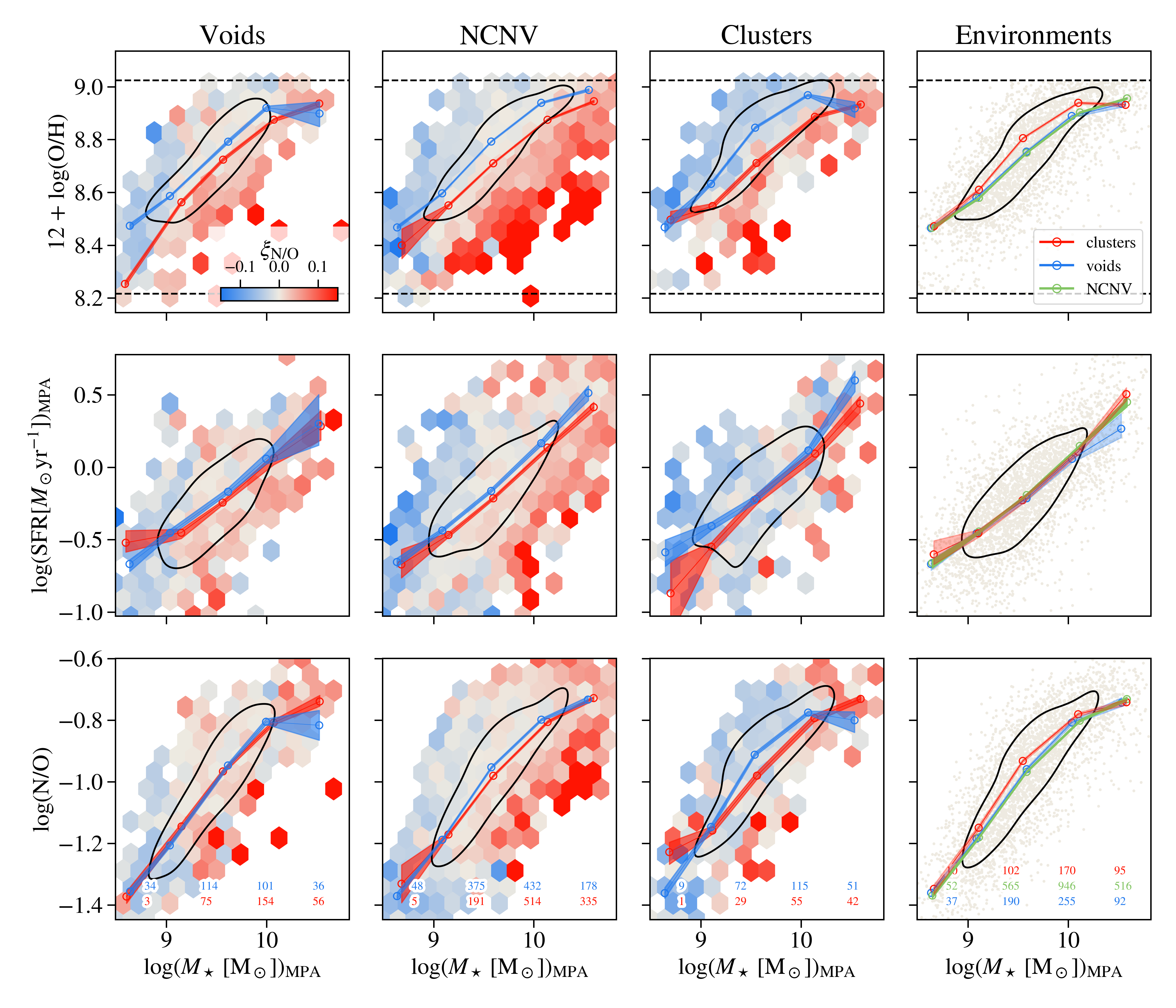}
    \caption{
    Same as Fig.~\ref{fig:main_relations}, but also dividing into environmental subsamples.
    In the bottom right panel, the number of galaxies for different environments (clusters, NCNV, and voids) and each \mstar{} bin.}
    \label{fig:main_relations_environment}
\end{figure*}

In Sect.~\ref{subsec:candidate_params}, we suggested a connection between the MZR and the \nooh{} relation in which deviations from one scaling law find their counterpart in the other.
Consequently, we would expect a different MZR relation for cluster galaxies. To check this, we plot in Fig.~\ref{fig:main_relations_environment} the MZR, SFMS, and N/O-\mstar{} relation across different environments and further segregated into \excessNO{}$\lessgtr0$ subpopulations, mirroring the framework used in Fig.~\ref{fig:main_relations}. We find no differences between voids and NCNV, but a clear distinction for clusters. Cluster galaxies show higher O/H at fixed \mstar{}, and that difference is further emphasised by the division in \excessNO{}, as the general shape of the cluster sample MZR is driven by those galaxies with \excessNO{}$<0$. In fact, cluster galaxies with \excessNO{}$<0$ outnumber those with \excessNO{}$>0$ by a ratio of 2:1, as opposed to the ratio of 1:1 for voids and NCNV.
Moreover, cluster galaxies do not show significant differences in the SFR compared to NCNV and voids (Fig.~\ref{fig:main_relations_environment}, middle rightmost panel). This is expected based on the Spearman correlation analysis: differences in the SFR are not correlated with different \nooh{} relations in cluster galaxies. Again, these results suggest that deviations in the \nooh{} relation are driven by mass-related processes, such as mass assembly paths, or environmental processes rather than instantaneous observed SFR activity. Additionally, in that same figure, cluster galaxies show a slight positive offset also in the N/O-\mstar{} relation, where they exhibit higher N/O ratios at fixed mass than either NCNV or voids.

These results are consistent with the small offsets in the \nooh{} relation (up to $\sim 0.05$~dex  lower $\text{N/O}$ or higher $\text{O/H}$) reported by \citet{Pilyugin2002} for $\ion{H}{II}$~regions in Virgo cluster spirals, with respect to a field control sample. However, while \citet{Pilyugin2002} cautioned that their findings might be artefacts of extrapolating the $P$-method \citep{Pilyugin2001} into a regime of near-zero excitation ($P = R_3 / R_{23}$) and low $R_{23}$  values  ($R_{23} < 1$), our analysis rules out this uncertainty. In our sample, although the $P$ distribution at the high-metallicity end is slightly skewed toward lower values for cluster galaxies, it remains fully compatible with the baseline range observed in NCNV. Furthermore, the \citet{Mcgaugh1991} $R_{23}$ calibration used here relies on photoionisation models that explicitly account for ionisation parameter variations via  $\rm{O}_{32}$. The underlying model grid is valid over a wide parameter space ($R_{23}> 0.1$ and spanning an excitation range, which in terms of $P$, comprises the interval $0.03 \leq P \leq 0.97$), safely encompassing our entire dataset, and ensuring that our abundance derivations are not affected by extrapolation biases.

Although direct studies of the \nooh{} relation within the context of the LSS are scarce, the number of works focused on the MZR is considerably more extensive. Following this, and taking into account the previously stated connection between the scatter in the MZR and the \nooh{} relation, we can compare our findings with literature results. 

To date, observational studies \citep[e.g.][]{Ellison2009, Hughes2013} and simulations \citep[e.g.][]{Rowntree2024} remain inconclusive as to whether systematic MZR variations exist across different LSS environments. Furthermore, when such discrepancies are reported, they are typically confined to the highest-density environments, such as clusters. In particular, 
\citet{Ellison2009} report, on average, higher metallicities by up to 0.04 dex in cluster galaxies. The authors suggest that the observed chemical enhancement in clusters may be driven by local overdensity rather than by cluster membership. Consistent with this, several studies have reported that galaxies in extremely dense environments, defined using a local-density approach, exhibit an increase in O/H relative to their lower-density counterparts~\citep{Cooper2008, Peng2014, Wu2017}. Nonetheless, which mechanism is the main driver for these differences remains an open question. Some of the proposed hypotheses include strangulation~\citep[e.g.][]{Peng2015}, and metal-rich gas inflows~\citep[e.g.][]{Peng2014}.

\citet{Petropoulou2012} found that the MZR and N/O$-$\mstar{} relations for low-mass galaxies ($\log M_\star[M_\odot] \in [8, 10]$) in massive clusters exhibit a positive offset in both the N/O-\mstar{} relation and MZR when compared to other less massive clusters. 
Specifically, they find higher N/O and O/H values at fixed \mstar{} (despite using nitrogen-based O/H calibrations) consistent with the differences we identify when comparing cluster galaxies against NCNV and void populations.
Among the hypotheses presented in that work, we highlight those related to gas outflows and inflows (and suggest that the reader check the thorough discussion presented there). In an `equilibrium' model in which the MZR is the consequence of a competing balance between gas inflow and outflows (e.g. differential galactic winds), the presence of an intracluster medium (ICM) could lead to two distinct effects that can produce the observed trends: the suppression of infall of pristine gas and more efficient re-accretion of enriched gas. The former scenario would imply that a significant fraction of the SF galaxies in clusters would consume their gas with a reduced gas supply (a time-extended quenching or strangulation), thus presenting an abnormally high O/H before becoming quenched. The latter phenomenon could, in turn, be caused by the high ambient pressure of the ICM, which reduces the outflow of metal-rich gas. Both effects would lead to a higher O/H at fixed \mstar{} when compared to galaxies in other low-density environments. Thus, both scenarios are plausible and would explain our findings. Interestingly, although both an increase in N/O and O/H take place at fixed \mstar{} for cluster galaxies (see again Fig.~\ref{fig:main_relations_environment}), the net effect is a decrease in \excessNO{} when comparing the \nooh{} relation, as the increase in O/H is much larger than that of N/O. As another explanation for these differences in clusters (causing also an increase in O/H at fixed \mstar{}), ~\citet{Khoram2025} pointed to ram pressure stripping as the main cause behind differences in the MZR, although this effect would be more significant at higher redshift.

Focusing on low-density environments, \citet{Molina-Calzada2025} found that void galaxies present lower O/H and higher SFR at fixed \mstar{} than galaxies in filaments or tendrils, a difference that is not statistically significant in our analysis. However, their methodology is not directly comparable to ours, as they make use of a \mstar{}$-(g-r)$ controlled `field' sample for comparison with each of their environmental samples. For the sake of comparison, we constructed a similar control sample where void, NCNV, and cluster galaxies are matched in the \mstar{}$-(g-r)$ distributions up to $0.02$~dex and $0.06$~mag, respectively. Even with this controlled sample, we find no significant differences either in the MZR, SFMS, or $\excessNO{}-\tvox{}$ relations. Although we cannot replicate the trends in~\cite{Molina-Calzada2025}, this discrepancy is likely driven by a combination of limited sample size (in both works), selection effects, and methodological differences: we highlight the broader redshift range $z<0.3$, a different SF definition, and the use of different O/H abundance estimators (in their case, they employ O3N2, a nitrogen-based index; see Sect.~\ref{subsec:chemical} and references therein for a discussion on the topic). In fact,~\citet{Conrado2026} also find no statistically significant differences in the integrated SFR properties between voids and a morphology-controlled `filament' sample (differences only appear when using resolved information).

In the dwarf galaxy regime, conflicting results have been found with either no significant differences for void galaxies~\citep{Kreckel2015}, or only marginal ones~\citep{Pustilnik2011}, although both studies were limited by small sample sizes (10 and 48 galaxies, respectively). 
More recently, \citet{Bidaran2026} analysed a large sample of SDSS dwarf galaxies ($\log \mstar{}\in[8.9, 9.5])$ and found that cluster galaxies show systematically higher \tvox{} at fixed \mstar{} than those in voids or NCNV, a trend that closely mirrors our own findings.
They find no statistically significant differences in the MZR when comparing void and NCNV galaxies (see also Table B.1 therein for $R_{23}$-derived O/H results). The compatibility of their results with ours is to be expected, given that both studies utilise galaxy samples extracted from the same SDSS parent sample.

In Appendix~\ref{app:biases}, we again briefly address the potential biases introduced by the S/N threshold, differences in the $z$ distribution, and aperture effects on the comparison of the \nooh{} relation between different LSS environments, as well as comment on other factors that might be in place (and cannot be tested). The impact of the adopted chemical abundance calibrations on these results is described in Appendix~\ref{app:oxygen}. Lastly, we note that the reported systematic differences between environments are, strictly speaking, differences in the $R_{23}-\rm{N2O2}-O_{32}$ relations (transformed by the calibrations) and are highly determined by the underlying $R_{23}-\rm{N2O2}$ relation.

\section{Summary and conclusions}\label{sec:conclusions}
In this work, we study the impact of various galaxy properties on the scatter of the gas-phase \nooh{} relation and systematically analyse this relation across different LSS environments: clusters, NCNV, and voids. 

First, we identify the physical properties that best correlate with the vertical deviations ($\xi_\text{N/O}$) from the principal \nooh{} relation. Our findings indicate that \mstar{}-related variables, such as total \mstar{and $g-r$ colour, are the main drivers of the scatter. 
We interpret this result in connection with the MZR and find that these trends are physically compatible with the mixed effects of the time-delayed nitrogen production by low- and intermediate-mass stars, pristine gas inflows, and differential galactic winds. Nonetheless, our results do not allow us to discriminate between the three mechanisms. Although these trends can be partly modulated by methodological effects, such as aperture coverage and sample selection bias, a series of different robustness tests confirms the presence of an underlying dominant physical mechanism.

When evaluating environmental differences, we find that cluster galaxies exhibit systematically lower N/O than either NCNV or voids in the high-metallicity range (\tvox{} > 8.63; reaching deviations of up to \excessNO{} $\sim- 0.02$). In contrast, void galaxies show no statistically significant deviations from the NCNV baseline. Furthermore, these differences are also connected to a different MZR and N/O-\mstar{} relation for cluster galaxies, showing higher N/O and O/H values at fixed stellar mass.  
We briefly contextualise our results within the framework of environmental evolution, in which the presence of dense ICM can induce strangulation and favour the re-accretion of metal-enriched gas. 

This study demonstrates that the impact of the LSS on the \nooh{} relation is restricted to the highest-density regions of the Universe, namely galaxy clusters. To fully disentangle the effects of the cosmic web, future studies will require spatially resolved spectroscopy combined with galaxy samples defined homogeneously across LSS environments.

\begin{acknowledgements}

We acknowledge support from PID2023-149578NB-I00 and PID2023-150178NB-I00   financed by MCIU/AEI/10.13039/501100011033, and by FEDER, UE. We also acknowledge support via grant FQM108, financed by the Junta de Andalucía (Spain) and from the Grant AST22-4.4, funded by Consejería de Universidad, Investigación e Innovación (Junta de Andalucía) and Gobierno de España and Unión Europea – NextGenerationEU.
YGK acknowledges financial support from PREP2023-001684 funded by MCIU/AEI/10.13039/501100011033 and the FSE+.
SFS acknowledges the support by CBF-2025-I-236 project granted by the Secretaría de Ciencia, Humanidades, Tecnología e Innovación (SECIHTI) of
the Mexican Federal Government, and the PID2022-136598NB-C31 (ESTALLIDOS) grant by the Spanish Ministry of Science and Innovation (MCINN). 
Funding for the Sloan Digital Sky 
Survey IV has been provided by the 
Alfred P. Sloan Foundation, the U.S. 
Department of Energy Office of 
Science, and the Participating 
Institutions. 

SDSS-IV acknowledges support and 
resources from the Center for High 
Performance Computing  at the 
University of Utah. The SDSS 
website is www.sdss4.org.

SDSS-IV is managed by the 
Astrophysical Research Consortium 
for the Participating Institutions 
of the SDSS Collaboration including 
the Brazilian Participation Group, 
the Carnegie Institution for Science, 
Carnegie Mellon University, Center for 
Astrophysics | Harvard \& 
Smithsonian, the Chilean Participation 
Group, the French Participation Group, 
Instituto de Astrof\'isica de 
Canarias, The Johns Hopkins 
University, Kavli Institute for the 
Physics and Mathematics of the 
Universe (IPMU) / University of 
Tokyo, the Korean Participation Group, 
Lawrence Berkeley National Laboratory, 
Leibniz Institut f\"ur Astrophysik 
Potsdam (AIP),  Max-Planck-Institut 
f\"ur Astronomie (MPIA Heidelberg), 
Max-Planck-Institut f\"ur 
Astrophysik (MPA Garching), 
Max-Planck-Institut f\"ur 
Extraterrestrische Physik (MPE), 
National Astronomical Observatories of 
China, New Mexico State University, 
New York University, University of 
Notre Dame, Observat\'ario 
Nacional / MCTI, The Ohio State 
University, Pennsylvania State 
University, Shanghai 
Astronomical Observatory, United 
Kingdom Participation Group, 
Universidad Nacional Aut\'onoma 
de M\'exico, University of Arizona, 
University of Colorado Boulder, 
University of Oxford, University of 
Portsmouth, University of Utah, 
University of Virginia, University 
of Washington, University of 
Wisconsin, Vanderbilt University, 
and Yale University.

This work made use of the following software packages: \texttt{astropy} \citep{astropy:2013,astropy:2018,astropy:2022,astropy_21262391}, \texttt{matplotlib} \citep{Hunter:2007}, \texttt{numpy} \citep{numpy}, \texttt{pandas} \citep{mckinney-proc-scipy-2010,pandas_21500199}, \texttt{python} \citep{python}, and \texttt{scipy} \citep{2020SciPy-NMeth,scipy_20383145}.

This research has made use of the Astrophysics Data System, funded by NASA under Cooperative Agreement 80NSSC21M00561.

Software citation information aggregated using \texttt{\href{https://www.tomwagg.com/software-citation-station/}{The Software Citation Station}} \citep{software-citation-station-paper,software-citation-station-zenodo}.
\end{acknowledgements}

\bibpunct{(}{)}{;}{a}{}{,} 
\bibliographystyle{aa} 
\bibliography{biblio}

\appendix
\section{Full spectral fitting algorithm}\label{sec:appendix_pyfit3d}

The full-spectral analysis is performed with a modified version of the pyFIT3D \texttt{python} module~\citep{Lacerda2022}. Some changes were made to the code that can be broadly categorised into (i) bug fixes, (ii) improvements in uncertainty estimation, and (iii) computing time optimisation. Regarding (i), the most relevant change affects redshift estimation, which was on average overestimated by $15$ km/s (with some extreme cases up to $45$~km/s). With respect to (ii), the errors associated with the single stellar population (SSP) coefficients, as well as those for the light- and mass-weighted mean ages and metallicities, were re-estimated as the standard deviation derived from the internal Monte Carlo iterations, obtained by resampling the spectra within the flux density error bars. This approach substitutes the previous approximation, which relied on a linear expansion, with the aim of providing more realistic uncertainty estimates, as some values were significantly underestimated. 
The most relevant changes affecting (iii) correspond to the optimisation of both the determination of redshift and the internal extinction. In the present version of the code, the former step is performed by displacing the observed spectrum to the rest frame rather than transforming all the models to the observed frame. The extinction is calculated by first joining together spectral channels in integrated bands of 5 \AA~\citep[in the same way as][but with smaller bands]{Sanchez2025}. All in all, the time execution reduction factor ranges from $2$ to $10$ in the best-case scenarios, depending on the input parameters. For an in-depth description of the changes, see González-Koda et al. (in prep.), \url{gitlab.com/yllari/py-pipe-4-u}.

\section{\nooh{} relation of composite galaxies}\label{app:composite}
As mentioned in Sect.~\ref{sect:Methodology}, composite galaxies were excluded from our main analysis. Nonetheless, since nitrogen and oxygen emission lines are used in the SF classification, and given our focus on the related N/O abundance ratio, we briefly discuss the implications of imposing different demarcation lines in the $[\ion{N}{II}]$ diagnostic diagram to define the galaxy sample: specifically, those of ~\citet{Kewley2001} and~\citet{Kauffman2003a}. For this, we visually compare the results of the two classifications in the complementary $[\ion{S}{II}]$ and $[\ion{O}{I}]$ diagnostic diagrams.

In Fig.~\ref{fig:diagnostic}, we show the location of the galaxies in the three diagrams, where the solid lines correspond to the demarcation line proposed by~\citet{Kewley2001} and the dashed line in the leftmost panel corresponds to the demarcation line proposed by~\citet{Kauffman2003a}. There is a general agreement in the classification of galaxies as SF using Kauffman's criterion (grey points) and the lines proposed in~\citet{Kewley2001} in the $[\ion{S}{II}]$ and $[\ion{O}{I}]$ diagrams, despite the higher scatter in the  $[\ion{O}{I}]$ diagram due to the lower S/N ratio of this line. The same conclusions apply to the AGN classification given by~\citet{Kewley2001} in the $[\ion{N}{II}]$ diagram (orange points). Regarding those galaxies that fall between ~\citet{Kewley2001}'s and ~\citet{Kauffman2003a}'s curves (in green, Fig~\ref{fig:diagnostic}), usually referred to as `composite' objects,  it is clear that, while they would be classified as non-SF by the stricter \citet{Kauffman2003a}'s criterion, they lie well below the SFR regime defined in both the $\collisional{S}{II}{6717}$ and  $\collisional{O}{I}{6300}$ diagrams. Under a similar analysis,~\citet{Perez-Montero2009}, argued that these objects could be associated with high-N/O SF galaxies with peculiar chemical evolution histories.  Thus, discarding them entirely could introduce systematic selection biases.

\begin{figure}
    \centering
    \includegraphics[width=\linewidth]{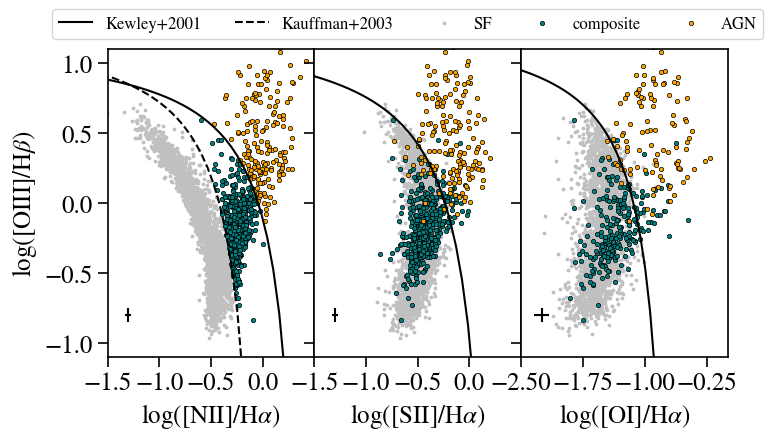}
    \caption{From left to right, $\collisional{N}{II}{6583}$, $\collisional{S}{II}{6717}$, and $\collisional{O}{I}{6300}$ diagnostic diagrams for the whole sample. The dashed and solid black lines indicate~\citet{Kauffman2003a} and~\citet{Kewley2001}'s demarcation lines. In grey, green and orange, the data points indicate whether the galaxies are classified as SF by~\citet{Kauffman2003}'s demarcation line, composite or AGNs, respectively. In the bottom left corner of each panel, we show the median error bars.}
    \label{fig:diagnostic}
\end{figure}

As shown in Fig.~\ref{fig:composite}, when we translate the composite population onto the \nooh{} plane, these galaxies follow a sequence that lies systematically above the main SF one. This deviation is consistent with the analysis performed in Sect.~\ref{subsec:candidate_params}, as composite galaxies are generally redder, more massive and exhibit lower sSFR than their pure SF counterparts~\citep[e.g][]{Wu2022}. These observed properties are compatible with both a non-SF contribution to the emission lines (such as AGN and diffuse ionised gas emission) and the presence of SF embedded within an older stellar population, making it hard to disentangle the two effects. For example, the position of composite galaxies in the \nooh{} plane would be in line with the periodic gas infall model proposed by~\citet{Koppen2005}. 

Since a proper treatment of this population of galaxies would require a separate in-depth study, we excluded them to limit the scope of this work. However, we have tested that including them in our galaxy sample results in the same qualitative results. The effects of including the composite population are: higher Spearman correlation coefficients in Fig.~\ref{fig:ssfr_indic} and larger separations between the positive and negative \excessNO{} in Fig.~\ref{fig:main_relations}.

\begin{figure}[htpb]
    \centering
    \includegraphics[width=
    \linewidth]{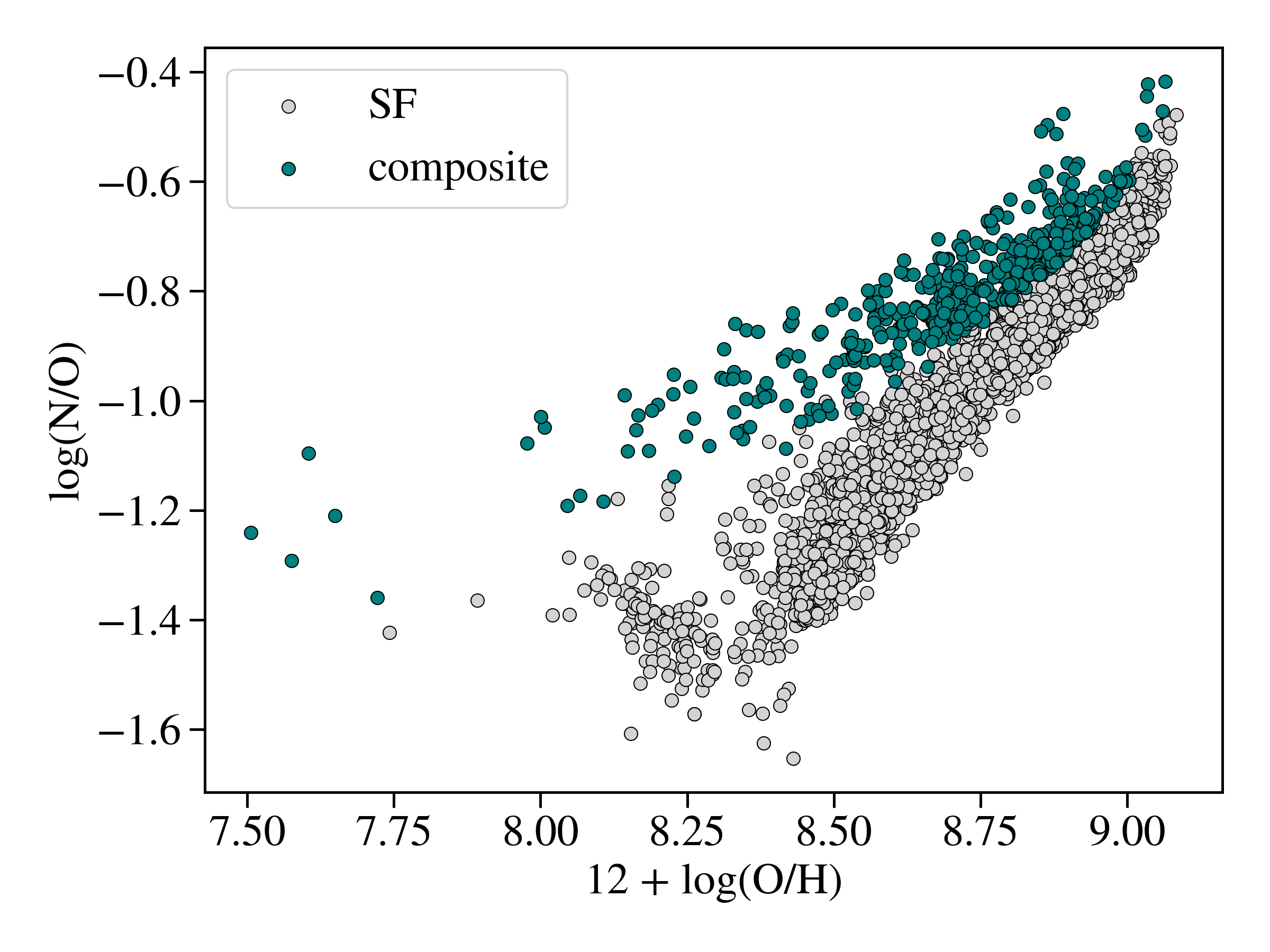}
    \caption{N/O versus O/H for SF galaxies as considered in the main analysis (grey) and composite (teal) galaxies as given by~\citet{Kauffman2003a} and~\citet{Kewley2001}'s demarcation lines. }
    \label{fig:composite}
\end{figure}

\section{Other oxygen calibrations}\label{app:oxygen}
We dedicate this appendix to important information about all other chemical abundances explored and not included in Sect.~\ref{sect:results}. To this end, we provide the analytical expression used in this work for~\citet{Curti2017,Curti2020}'s $\log($O3S2) and $\log(R_{23})$, shortly summarise the methodology followed for the~\citep{Garner2025} photoionisation grid scheme (photogrid, hereafter), as well as explore the consistency of our results when using different calibrations of $R_{23}$ to derive the oxygen abundance. 

In ~\citet{Curti2017,Curti2020}, the relations provided are expressed as $\log R=\sum_{n}c_nx^n$, where $R$ is the corresponding index. To obtain an approximate inverse, we fit the function 
\begin{equation}
    12 + \log(\text{O/H}) = a + bx + c\sqrt{d -  x} + ex^2 ,
\end{equation}
using a least squares fitting, where $x\equiv \log R$. We restrict this fit to the monotonic ranges provided in~\citet{Curti2017,Curti2020}, namely $12 + \log \text{(O/H)}\in[7.6, 8.9]$ and $12 + \log \text{(O/H)}\in[8.4, 8.85]$ for the O3S2  and $R_{23}$ indices, respectively. The values $(7.962, 0.115, 1.008, 0.962, 0.)_{R_{23}}$ and $(6.974, 0.167, 1.167, 2.268,0.022)_\text{O3S2}$ provide a good fit with absolute residuals relative to the original functions bounded by 0.0005 and 0.006 within the given range, values significantly smaller than the typical observational uncertainties in O3S2 and $R_{23}$.

With respect to the photogrid derivation of the chemical abundances, the CLOUDY v23.01~\citep{Chatzikos2023} photoionisation models are used in conjunction with the NebulaBayes~\citep{Thomas2018} inference code to obtain posterior distributions for the N/O, O/H, and ionisation parameter, $ \mathcal{U}$ of the interstellar medium. To do so, the line fluxes are provided and a uniform prior constrained by the line ratios $R_{23}, \rm{O}_{23}$ and N2O2 is adopted. The double degeneracy of $R_{23}-\rm{O/H}$ is broken by introducing a Gaussian prior $\mathcal{N}(9, 0.5)$ that favours the metal-rich branch, as we expect most of the sample to lie on $\tvox{}>8.4$. For more details, see~\citet{Garner2025}.

\begin{figure}[htpb]
    \centering
    \includegraphics[width=\linewidth]{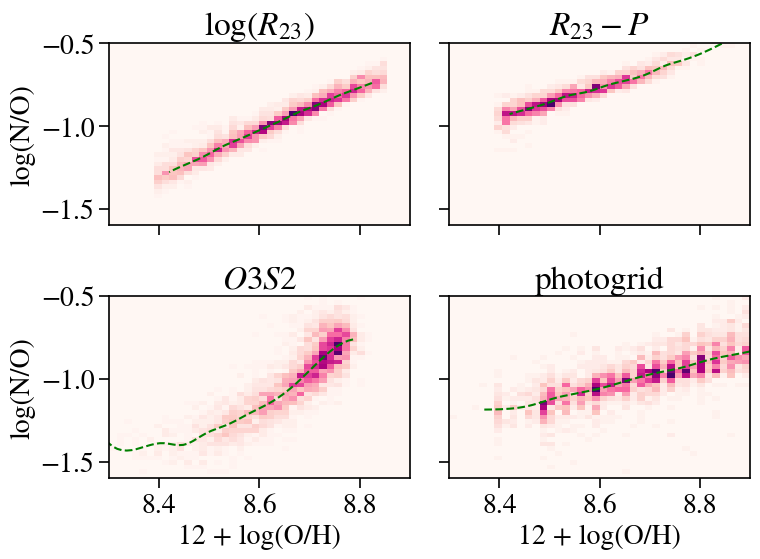}
    \caption{Distribution of galaxies in the \nooh{} plane for different calibrations presented in Table.~\ref{tab:calibrations}, where `photogrid', stands for the simultaneous Bayesian approach based on photoionisation grids. The distributions have been fitted with a spline function (green dashed line; see Sect.~\ref{subsec:candidate_params} for details). }
    \label{fig:app_diff_calibrs}
\end{figure}

In Fig.~\ref{fig:app_diff_calibrs}, we present the resulting \nooh{} relations for the different calibrations along with their respective spline fitting. A clear `knee' is visible in the  O3S2 distribution (bottom left panel) at the high O/H regime, a feature not observed in Fig.~\ref{fig:excess_definitions}, or when using $\log(R_{23})$, $R_{23}-P$, or `photogrid'. This suggests a systematic discrepancy inherent to the O3S2 calibration. Because of this, and as our methodology requires an invertible \nooh{} relation, we exclude O3S2 from the analysis.

We repeated the same analysis described in Sect.~\ref{subsec:candidate_params} (correlation of \excessNO{} with galaxy properties) and Sect.~\ref{subsec:no_oh_envs} (influence of the environment) by employing two alternative $R_{23}$ calibrations and found extremely similar results. To avoid redundancy, these findings are summarised in Fig.~\ref{fig:app_a_summary}, which compares the Spearman correlation coefficients for both the total stellar mass and the total SFR from MPA across the different calibrations of $R_{23}$ using both the observed and the mock data (from the test in Appendix ~\ref{sec:app_pearson}). The correlations for \mstar{} are compatible for all calibrations, with only slight differences. In the case of the SFR, the $R_{23}-P$ method exhibits slightly stronger correlations ($\rho\sim 0.2$). In the case of photogrid, we find that the resulting \nooh{} relation exhibits a significantly larger scatter compared to the other three calibrations (see also Fig.~\ref{fig:app_diff_calibrs}). This large scatter (and its correlation with the galaxy parameters) positions both the \mstar{} and SFR Spearman correlations under the baseline established by our mock sample, thus not allowing us to ascertain statistical significance in this particular case.
This suggests that, with this method, the physical signal is likely superseded by observational uncertainties, preventing us from establishing statistical significance for these parameters in this particular case. Additionally, Fig.~\ref{fig:app_a_summary} shows the difference in the median \excessNO{} values for clusters and voids (magnified $\times 10$ for visual clarity) relative to the NCNV baseline. Consistent with our main results, there is no significant difference for voids, while cluster galaxies systematically exhibit a negative offset of $\gtrsim0.02$.  This is true for every calibration and \nooh{} derivation.

\begin{figure}
    \centering
    \includegraphics[width=\linewidth]{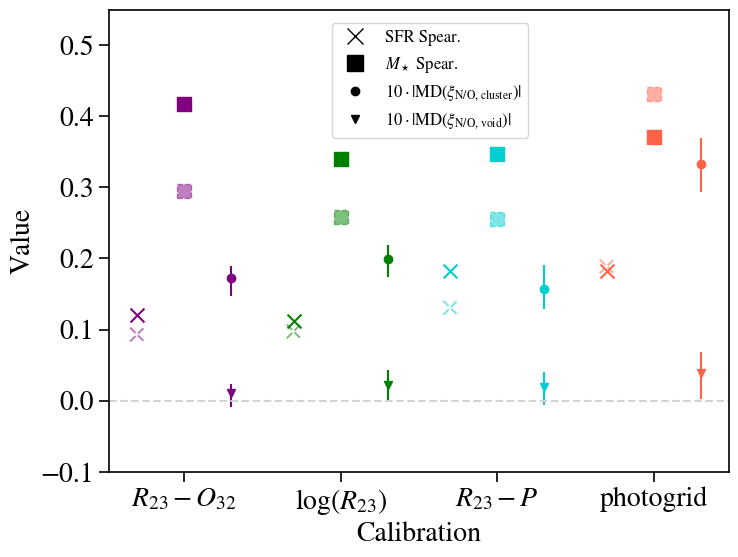}
    \caption{Effect of the alternative $R_{23}$ calibrations used for the derivation of O/H (Table~\ref{tab:calibrations}) on the obtained correlation of residuals and environmental offsets.  
    The plot shows the Spearman correlation coefficients ($\rho$) for total stellar mass (\mstar{}, squares) and star formation rate (SFR, crosses) across the different $R_{23}$ calibrations. Solid symbols indicate the observed data, while the shaded symbols represent results for their respective mock distributions. The calibration used in the main analysis (Sect.~\ref{sect:results}) is labelled as $R23-O32$. Additionally, the circles and inverted triangles display the median differences in \excessNO{} values with respect to the NCNV baseline (MD), for cluster and void galaxies, respectively. These differences have been magnified by a factor of 10 for visual clarity.}
    \label{fig:app_a_summary}
\end{figure}

\section{Spearman correlation tests}\label{sec:app_pearson}

This appendix explores the robustness of the methodology introduced in Sect.~\ref{subsec:candidate_params}. We present two tests designed to quantify the methodology's sensitivity to both observational errors and methodological assumptions: A mock test to quantify the impact of observational uncertainties and cross-correlation on the Spearman coefficient, and a test of the methodology in the case scenario in which the residual variables $\delta x$ are defined with respect to \mstar{} instead of \tvox{}.  The latter considers the effects of well-known \mstar{} relations, such as the MZR or the SFMS, on the \excessNO{}.

We first examine the scenario presented in Appendix~\ref{app:biases} in which no physical correlation exists between the candidate parameters and the \nooh{} scatter. To achieve this, we eliminate any correlation between the two variables (O/H and N/O) by randomising the O/H distance to the relation, $\delta[12 + \log (\text{O/H})]_i$, as given by the distance to the spline-fitting applied to the inverse relation O/H-N/O. That is, our mock sample consists of galaxies following the interpolated median relation to which we add uncorrelated noise in the O/H axis, thus creating the scatter. The distance to the main relation, i.e. the noise, is extracted from the pool of distances of the original sample to the interpolated relation and added to each data point\footnote{To account for variations in the scatter distribution along the main \nooh{} sequence, we bin the N/O axis and extract the distance from that `nearby points' pool.}. This procedure yields a sample with the exact same global $\delta \text{[\tvox{}]}$ distribution as the original one but without any underlying physical correlations.

We repeat the same steps followed in Sect.~\ref{sect:results}: the \nooh{} relation interpolation and \excessNO{}\ definition and candidate parameter analysis, but now for this mock sample. The resulting distribution of the Spearman coefficients is shown in Fig.~\ref{fig:ssfr_indic} as dashed empty violins. We conclude (see Sect.~\ref{subsec:candidate_params}) that the correlation coming from SFR-related variables could be completely explained by this scenario, but not those variables that are related to \mstar{}.

Lastly,  we re-evaluate our findings by defining the residual of each variable with respect to \mstar{}, rather than O/H. That is, the residual is now calculated as the difference with respect to the interpolated $x-M_\star$ relation. We present the results in Fig.~\ref{fig:mass_ssfr_indic}. As in the main analysis, for the reference variable (in this case, \mstar{}), the Spearman correlation is calculated using the variable itself rather than its residual. The results support the idea that the position of a galaxy relative to the O/H-\mstar{} relation correlates with its position in the \nooh{} relation. This is evidenced by the fact that the largest Spearman coefficients (in absolute value) correspond to \mstar{}, and $\delta$[\tvox{}]. Moreover, the deviations in the SFMS~\citep[see e.g.][]{Noeske2007, Renzini2015} show a weaker correlation with the \excessNO{}, as the correlation of the $\delta$ SFR $=$ $\delta$sSFR variables\footnote{By definition, the vertical distance to the SFMS and the vertical distance to the sSFR-\mstar{} relation are equal.} remains weaker than that of \mstar{} and O/H. This was also the case in the main analysis.

\begin{figure*}
    \centering
    \includegraphics[width=\linewidth]{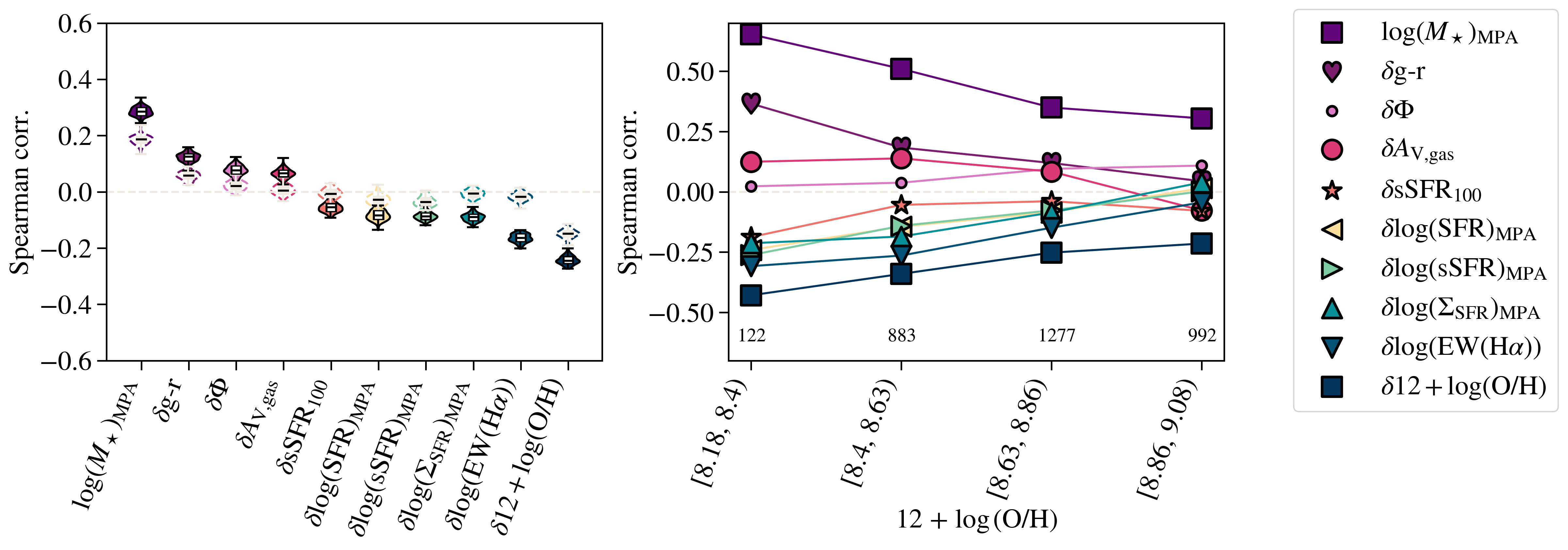}
    \caption{Same as Fig.~\ref{fig:ssfr_indic} of the main text but with mass as reference value to define the residual variables, $\delta x$.}
    \label{fig:mass_ssfr_indic}
\end{figure*}

\section{Assessing potential systematic biases}\label{app:biases}
First, we address whether the correlations found between the scatter in the \nooh{} relation and the candidate parameters could be artefacts of a biased definition of $\xi_\text{N/O}$. For example, if the cubic spline-fitting yielded a $\xi_\text{N/O}$ distribution that was off-centred at high O/H, it would introduce a spurious correlation between the explanatory variable and the residuals $\xi_\text{N/O}$. To test this, we included \tvox{} among the candidate variables (without applying any spline-fitting subtraction). As shown in the left panel of Fig.~\ref{fig:ssfr_indic}, the resulting distribution for the Spearman coefficient for the \tvox{}$-\xi_\text{N/O}$ relation is approximately centred at zero, confirming that no such systematic offset is being introduced by our methodology\footnote{Nonetheless, we further verified this by repeating the analysis using the absolute values instead of using the  `excess' (or residual) definition; in every instance, the results remain fully consistent with only minor variations in the parameter ranking and reduced correlation strengths.}. 

Next, we examine an extreme scenario where the observed correlations could emerge without a direct `physical' link: a case in which the \nooh{} relation is intrinsically tight, and the observed scatter is entirely caused by observational errors. In this scenario, mass-related variables would act as proxies for the true O/H (or N/O) due to the MZR and N/O-\mstar{} relation, which would convey information about the direction and magnitude of the observational deviations of the galaxy from the primary \nooh{} relation. Consequently, a spurious correlation emerges as the N/O residuals scale directly with the magnitude of the observational errors in O/H, creating an artificial dependency between the variables. 
Although this case is extreme, it can provide an upper bound to the potential bias introduced by such cross-correlations. This, along with other tests, is presented in Appendix~\ref{sec:app_pearson}, where we build a mock galaxy sample designed to have zero intrinsic physical correlation among the residuals. The resulting Spearman coefficients from this test are shown in the left panel of Fig.~\ref{fig:ssfr_indic} as dashed violin plots and grey box plots. These distributions represent the expected correlations arising from decorrelated observational scatter. Accordingly, we emphasise the significance of the highly correlated variables (\mstar{} and colour), as they clearly stand out from these mock-defined bounds,  in contrast to variables like SFR, which are consistent with the correlations expected from observational scatter alone.

Finally, we study other potential biases in our methodology: the S/N criteria effects, the aperture coverage and the choice of oxygen abundance calibration.
Regarding the S/N, we have checked that the results remain consistent regardless of the \texttt{snMedian} threshold or the S/N limit imposed on the emission-line fluxes. To address aperture effects, we have confirmed that the mass correlation is present and strong via an alternative test: We divide the sample into two high- and low-mass subsamples (defined by imposing $\log(M_\star) \lessgtr \text{median}(\log(M_\star))$) matched in their redshift (also tested on the Petrosian radius 90 distributions). Next, we calculate the difference between the medians of their respective \excessNO{} distributions. The test results in a significant difference between the two distributions and their medians of $d_f\sim 0.021$~dex. Applying the same test to the mock baseline from Appendix~\ref{sec:app_pearson} yields a lower median difference of $d_f' \sim 0.01$~dex. To ensure that the differences are statistically significant, we create 300 mock samples and calculate in every instance if $d_f > d_f'$. We obtain this to be true for every mock sample, bounding the effective p-value $p<1/300$ for the alternative hypothesis\footnote{This effective p-value is only an indication of whether the mock scenario could explain the differences. The way in which this hypothesis reflects the statistical significance of the results is completely determined by this assumption.} $d_f > d_f'$. This simple test rules out the possibility that the joint effect of differences in the aperture coverage and statistical cross-correlation drives the found correlations. Lastly, we find that our conclusions are robust across different oxygen calibrations and the simultaneous N/O and O/H determination introduced in Sect.~\ref{subsec:chemical} (see Appendix~\ref{app:oxygen}).

As a final remark on the methodology, we would like to note that the conclusions presented here have been demonstrated to be robust in describing the galaxy properties driving the observed trends in the $\log(R_{23})-\log(\rm{O_{32}})-\log(\rm{N2O2})$ space for integrated measurements. Naturally, these results should be understood in the context of single-fiber observations, encompassing emission from multiple $\ion{H}{II}$ regions alongside potential minor contributions from secondary ionisation mechanisms ---such as shocks, diffuse ionised gas, post-asymptotic giant branch stars, or AGNs---~\citep[e.g.][]{Zhang2017, Sanders2017}, although our strict star-forming selection criteria (Sect.~\ref{sect:Methodology}) ensure that the sample is overwhelmingly dominated by stellar photoionisation. Within this framework, the detected abundance variations reflect physical trends in galaxy properties, although they may include contributions that we cannot fully quantify with single-fibre data alone due to: a) differences in the global distribution and properties of $\ion{H}{II}$ regions within the aperture, and b) subtle variations in the ionisation conditions across the integrated star-forming complexes.

Regarding the comparison of the \nooh{} relation between LSS environments, we also address the potential biases introduced by the S/N threshold, differences in the $z$ distribution, and aperture effects. With respect to the S/N threshold, we have verified that all results remain consistent when considering \texttt{snMedian} > 5, 10, 20 and different S/N cuts for the emission-line fluxes. Concerning $z$, we have checked that there are no significant changes in the results when restricting the sample to a narrower redshift range ($z<0.04$) to match the redshift distribution across environments. Regarding the aperture effects, matching the three samples to have the same $\log M_\star-z$ distribution (matched up to $\varDelta\log(M_\star)=0.06$ dex and $\varDelta z=0.0005$, respectively; corresponding to a $50\times 50$ grid), does not alter our findings, although it significantly reduces the size of the sample. Additionally, we note that~\citet{Perez2025} found that late-type cluster galaxies in the mass range $\log \mstar[M_\odot]\in [9, 10.5]$ tend to be smaller than their counterparts in voids and NCNV ~\citep[although ][does not find any differences]{Abdullah2026}. This effect could have an impact on the observed N/O and O/H relations, as the fibre coverage may be systematically probing different physical scales (relative to the effective radii) for cluster galaxies.  On the other hand, the distinctive behaviour of cluster galaxies could suggest differences in the relative contribution of diffuse ionised gas or other sources of ionisation to the spectrum. Overall, a deep understanding of this matter would require resolved data on a self-consistently defined LSS sample.


\end{document}